%% file: main.tex
\documentclass[9pt, conference]{IEEEtran}
\IEEEoverridecommandlockouts
\usepackage{cite}
\usepackage{amsmath,amssymb,amsfonts}
\usepackage{algorithmic}
\usepackage{graphicx}
\usepackage{textcomp}
\usepackage{xcolor}
\usepackage{tcolorbox}
\usepackage{listings}
\usepackage{inconsolata}
\usepackage{multirow}
\usepackage{threeparttable}
\usepackage{booktabs}
\usepackage{pifont}
\usepackage{bbding}
\usepackage{colortbl}
\usepackage{subcaption}
\usepackage{wrapfig}
\usepackage{url}

\newcommand{\ours}{\emph{VeriPrefer}}
\newcommand{\oursft}{\emph{VeriPrefer}$_{\rm \bf SFT}$}
\newcommand{\oursrl}{\emph{VeriPrefer}$_{\rm \bf RL}$}
\newcommand{\x}{\boldsymbol{x}}
\newcommand{\y}{\boldsymbol{y}}
\newcommand{\yplus}{\boldsymbol{y}^+}
\newcommand{\yminus}{\boldsymbol{y}^-}
\newcommand{\yhat}{\hat{\boldsymbol{y}}}
\newcommand{\dsft}{\mathbb{D}_{\rm \bf SFT}}
\newcommand{\drl}{\mathbb{D}_{\rm \bf RL}}
\newcommand{\pass}{\emph{pass@}}
\newcommand{\p}{\emph{p@}}

\newcommand{\plus}[1]{\textbf{($\raisebox{1pt}{$\scriptstyle\boldsymbol{+}$}$#1)}}
\newcommand{\minus}[1]{\textbf{($\raisebox{1pt}{$\scriptstyle\boldsymbol{-}$}$#1)}}

\definecolor{promptgray}{RGB}{49, 49, 49}
\newcommand{\prompt}[1]{%
  \raisebox{-0.2em}{%
    \tcbox[
      size=tight,
      colback=promptgray, 
      colframe=promptgray, 
      fontupper=\footnotesize\color{white}, 
      left=1mm,
      right=1mm,
      top=0.5mm,
      bottom=0.5mm,
      arc=3mm, 
      nobeforeafter,
    ]{#1}%
  }%
}

\definecolor{vcspink}{rgb}{1.0, 0.6, 0.7}
\newcommand{\vcs}[1]{%
  \setlength{\fboxsep}{1pt}%
  \setlength{\fboxrule}{1pt}%
  \fcolorbox{black}{vcspink}{\color{black}#1}%
}

\newcommand{\custombox}[1]{%
  \setlength{\fboxsep}{1.5pt}%
  \setlength{\fboxrule}{1pt}%
  \fcolorbox{black}{white}{\color{black}#1}%
}

\definecolor{gold}{RGB}{255,215,0}
\definecolor{silver}{RGB}{192,192,192}
\definecolor{bronze}{RGB}{205,158,130}

\definecolor{darkgreen}{rgb}{0,0.5,0}
\definecolor{keypink}{RGB}{204, 84, 205}
\definecolor{keybackblue}{RGB}{225, 232, 255}
\definecolor{keybackgreen}{RGB}{227, 234, 209}
\definecolor{keybackorange}{HTML}{FFD6A5}
\definecolor{keyblue}{HTML}{005FD7}
\definecolor{keygreen}{HTML}{005F00}
\definecolor{namepurple}{HTML}{5F00D7}
\definecolor{numberorange}{HTML}{D75F00}
\definecolor{llmblue}{RGB}{149, 206, 255}
\definecolor{vcspurple}{RGB}{207, 88, 238}
\lstdefinestyle{plain}{
    basicstyle=\fontsize{6}{6}\ttfamily,
    columns=fullflexible,
    keywordstyle=\color{blue},
    commentstyle=\color{gray},
    stringstyle=\color{green},
    showstringspaces=false,
    breaklines=true,
    breakatwhitespace=false,
    breakindent=0pt,
    escapeinside={(*@}{@*)},
    aboveskip=0pt,
    belowskip=0pt,
    xleftmargin=0pt,
}
\lstdefinestyle{verilog}{
    language=Verilog,
    basicstyle=\fontsize{6}{6}\ttfamily,
    keywordstyle=\color{keyblue},
    keywordstyle=[2]\color{black},
    commentstyle=\color{gray},
    stringstyle=\color{darkgreen},
    showstringspaces=false,
    breakatwhitespace=false,
    breaklines=true,
    breakindent=0pt,
    belowskip=0pt,
    escapeinside={(*@}{@*)},
    deletekeywords={reg, wire, always, begin, end},
    morekeywords=[2]{reg, wire},
    otherkeywords={always, begin},
    morekeywords=[3]{always, begin, end},
    keywordstyle=[3]\color{keypink},
    emphstyle={[1]\color{namepurple}},
    emph={[2]ns, ps},
    literate=
        {\'b}{{{\color{numberorange}'b}}}2
        {\'h}{{{\color{numberorange}'h}}}2
        {\'d}{{{\color{numberorange}'d}}}2
        {0}{{{\color{numberorange}0}}}1
        {1}{{{\color{numberorange}1}}}1
        {2}{{{\color{numberorange}2}}}1
        {3}{{{\color{numberorange}3}}}1
        {4}{{{\color{numberorange}4}}}1
        {5}{{{\color{numberorange}5}}}1
        {6}{{{\color{numberorange}6}}}1
        {7}{{{\color{numberorange}7}}}1
        {8}{{{\color{numberorange}8}}}1
        {9}{{{\color{numberorange}9}}}1
}

\lstdefinestyle{report}{
    language=Verilog,
    basicstyle=\fontsize{7}{7}\ttfamily,
    keywordstyle=\color{keyblue},
    keywordstyle=[2]\color{keygreen},
    commentstyle=\color{gray},
    stringstyle=\color{darkgreen},
    showstringspaces=false,
    breakatwhitespace=false,
    breaklines=true,
    breakindent=0pt,
    belowskip=0pt,
    escapeinside={(*@}{@*)},
    deletekeywords={reg, wire, always, begin, end},
    morekeywords=[2]{reg, wire},
    otherkeywords={always, begin},
    morekeywords=[3]{always, begin, end},
    keywordstyle=[3]\color{keypink},
    morekeywords={endmodule},
    emph={[1]serial_audio_encoder},
    emphstyle={[1]\color{namepurple}},
    emph={[2]ns, ps},
    emphstyle={[2]\color{keyblue}},
    emph={[3]is_i2s},
    emphstyle={[3]\color{black}},
    numbers=left,
    numberstyle={\scriptsize\color{black}\rmfamily},
    numberblanklines=true,
    numberfirstline=true,
    numbersep=4pt,
    xleftmargin=15pt,
    framexleftmargin=15pt,
    literate=
        {\'b}{{{\color{numberorange}'b}}}2
        {\'h}{{{\color{numberorange}'h}}}2
        {\'d}{{{\color{numberorange}'d}}}2
        {0}{{{\color{numberorange}0}}}1
        {1}{{{\color{numberorange}1}}}1
        {2}{{{\color{numberorange}2}}}1
        {3}{{{\color{numberorange}3}}}1
        {4}{{{\color{numberorange}4}}}1
        {5}{{{\color{numberorange}5}}}1
        {6}{{{\color{numberorange}6}}}1
        {7}{{{\color{numberorange}7}}}1
        {8}{{{\color{numberorange}8}}}1
        {9}{{{\color{numberorange}9}}}1
        {0/1}{{{\color{red}0/1}}}3
        {1/1}{{{\color{black}1/1}}}3
        {==>}{{{\color{red}==>}}}3
}

\usepackage{tikz}
\newcommand{\circledblack}[1]{\tikz[baseline=(char.base)]{
  \node[shape=circle, 
        draw, 
        fill=black, 
        text=white, 
        inner sep=1pt] (char) {#1};}}

\def\BibTeX{{\rm B\kern-.05em{\sc i\kern-.025em b}\kern-.08em
    T\kern-.1667em\lower.7ex\hbox{E}\kern-.125emX}}
\begin{document}

\title{
Insights from Verification: Training a Verilog Generation LLM with Reinforcement Learning with Testbench Feedback}

\author{
Ning Wang$^{1\dagger}$ \quad Bingkun Yao$^{1*\dagger}$ \quad Jie Zhou$^{2}$ \quad \quad Yuchen Hu$^{2}$ \quad Xi Wang$^{2}$ \quad Nan Guan$^{1}$ \quad Zhe Jiang$^{2}$ \\[0.5em]
$^{1}$City University of Hong Kong \quad $^{2}$Southeast University \\
$\dagger$ These authors contribute equally.
}

\maketitle

\begin{abstract}
Large language models (LLMs) have shown strong performance in Verilog generation from natural language description. However, ensuring the functional correctness of the generated code remains a significant challenge.
This paper introduces a method that integrates verification insights from testbench into the training of Verilog generation LLMs, aligning the training with the fundamental goal of hardware design: functional correctness.
The main obstacle in using LLMs for Verilog code generation is the lack of sufficient functional verification data, particularly testbenches paired with design specifications and code.
To address this problem, we introduce an automatic testbench generation pipeline that decomposes the process and uses feedback from the Verilog compiler simulator (VCS) to reduce hallucination and ensure correctness. We then use the testbench to evaluate the generated codes and collect them for further training, where verification insights are introduced.
Our method applies reinforcement learning (RL), specifically direct preference optimization (DPO), to align Verilog code generation with functional correctness by training preference pairs based on testbench outcomes.
In evaluations on VerilogEval-Machine, VerilogEval-Human, RTLLM v1.1, RTLLM v2, and VerilogEval v2, our approach consistently outperforms state-of-the-art baselines in generating functionally correct Verilog code.
We open source all training code, data, and models at \url{https://anonymous.4open.science/r/VeriPrefer-E88B}.
\end{abstract}


\section{Introduction}

Large language models (LLMs) have shown promising capabilities in various software programming tasks, which has led hardware design researchers to investigate their applications in various hardware design tasks. An task is to use LLMs for automatic generation of hardware description language (HDL) code, such as Verilog, from design specifications written in natural language.

Current works still aim to improve the functional correctness of open-source models. In this paper, we propose a method that incorporates verification insights provided by testbenches into the training of Verilog generation LLMs. Although the use of verification knowledge is not new and has been explored in previous work, such as OriGen~\cite{origen}, which applies self-reflection with compiler feedback to fix code and generate high-quality training data, our approach is the first to align LLMs with the fundamental goal of Verilog generation: functional correctness.


The main challenge in leveraging verification feedback for LLM training is the lack of sufficient functional verification data, specifically testbenches. Publicly available datasets rarely include detailed testbenches paired with design specifications and code, making it difficult to directly apply verification-based supervision. 
To address this limitation, we designed an automatic testbench generation pipeline given the Verilog code and design specification. This pipeline follows the principles of decomposition and feedback.
Decomposition breaks down the entire pipeline into smaller subtasks to prevent LLMs from being overwhelmed by task complexity. 
Feedback is provided through the verilog compiler simulator (VCS), a simulation and debugging tool developed by Synopsys for Verilog designs, which verifies correctness in time and provides different types of feedback to reduce the hallucination.
This pipeline collects verification insights and serves as a complementary source of knowledge from a different stage in the hardware design process. 
Although this pipeline cannot guarantee testbenches that cover all functional cases, it still supports training by providing basic functional verification.

\input{figs/simple_cap}
Our approach uses reinforcement learning (RL) to better align the Verilog generation model with functional correctness. We use RL because SFT only encourages the model to repeat patterns from the training data, which does not guarantee functional correctness. 
To address this, we prompt the fine-tuned LLM and collect generated code, then use testbenches to obtain preference pairs, where the preferred code passes more testcases and the less preferred code passes fewer.
We apply direct preference optimization (DPO), an RL algorithm that learns directly from these code pairs without modeling a reward function explicitly. This choice is motivated by the fact that reward-based RL is prone to reward hacking, where LLM exploits unintended shortcuts in the reward function \cite{rewardhacking}.
As a result, the model is not limited to repeating frequent patterns in the dataset, but is guided to produce outputs that better satisfy functional correctness. The choice of using DPO will be validated in Section \ref{sec:ablation} and Section \ref{sec:ppl}.

We evaluated our approach on three established benchmarks: VerilogEval-Machine, VerilogEval-Human and RTLLM v1.1, and found that our LLMs consistently outperform the state-of-the-art baselines of the general, coding and Verilog generation LLMs. The application of DPO leads to substantial improvements across all variants of the model with different structures, families, and sizes on all benchmarks. We further validate our approach on the advanced RTLLM v2 and VerilogEval v2 benchmarks, where the design specifications are more consistent with those used by HDL engineers, and our LLMs continue to outperform all state-of-the-art baselines. The gains achieved demonstrate that reinforcement learning with testbench feedback can better align the model with hardware design tasks to generate functionally correct Verilog code. 

The ablation study demonstrates that our approach achieves the best performance compared to various alternative strategies to construct preference pairs and RL algorithms. Furthermore, the ablation study shows that SFT with verified data does not achieve the same improvements as RL, since the training objective of SFT is perplexity, which measures the low-level cumulative token probability rather than the overall functional assessment of the entire code.

The contributions of this paper are as follows.
\begin{itemize}
    \item We propose an automatic testbench generation pipeline that decomposes the construction process and incorporates real verification feedback to ensure practicality and robustness.
    \item We integrate verification insights into the training process, addressing a key aspect of the hardware design task that is often neglected by previous work.
    \item We employ reinforcement learning with testbench feedback to better use verification insights, enabling the model to generate functionally correct Verilog code.
    \item We release all code, data, and the final model to promote further research and provide a strong baseline on established benchmarks.
\end{itemize}

\section{Related Work}
\subsubsection{Verilog Generation LLMs}
Many studies have advanced LLM-based Verilog code generation. \cite{rtlcoder} developed RTLCoder that outperforms GPT-3.5 by training on automatically generated datasets using GPT. \cite{betterv} introduced BetterV, which creates training datasets by converting Verilog code to the C language, allowing LLMs to use their knowledge of general-purpose programming languages. CodeV \cite{codev} used multilevel summarization, reversing the traditional process by generating natural language descriptions from high-quality Verilog code. Furthermore, AutoVCoder \cite{autovcoder} presented a systematic framework to improve Verilog code generation correctness through the generation of high-quality hardware datasets and the fine-tuning of two-round LLM. Furthermore, HaVen \cite{haven} addressed hallucinations in Verilog generation by classifying HDL-specific hallucinations and aligning LLM frameworks with engineering practices. OriGen \cite{origen} improved open-source LLM performance using self-reflection and code-to-code augmentation techniques. 
However, none of these approaches uses verification insights to train a Verilog generation LLM for better functional correctness.

\subsubsection{Training LLMs with Reinforcement Learning}
Recent advances in LLM have been significantly improved through RL techniques. The foundation of modern LLM alignment was established by Anthropic's work on reinforcement learning with human feedback (RLHF) \cite{rlhf}, which involves collecting human preference data to fine-tune models, balancing helpfulness and harmlessness objectives. Although effective, the complexity of the implementation of RLHF led to more efficient alternatives such as DPO \cite{dpo}, which enables direct optimization without explicit reward modeling.
In code generation, PPOCoder \cite{ppocoder} combines pre-trained models with proximal policy optimization (PPO) using compiler feedback as rewards, while IRCoCo \cite{ircoco} employs immediate rewards for code generation. PLUM \cite{plum} enhances the code language models by automatically generating test cases from natural language instructions and creating preference data by evaluating code solutions.
For Verilog generation, VeriSeek \cite{veriseek} represents the only work that uses code structure similarity as a reward to guide PPO training. However, this reward does not yet ensure that the generated code is functionally correct. Moreover, reward-based RL is prone to reward hacking, where LLM exploits unintended shortcuts in the reward function\cite{rewardhacking}.

\section{Method}
\ours{} accepts natural-language specifications as input and generates the corresponding Verilog code. \ours{} is developed on the foundation model through two sequential training stages. The first stage is SFT, which enhances the model's fundamental capability to respond to design specifications. 

In the second stage, we further align the model with functional correctness by incorporating verification feedback, as shown in Figure \ref{fig:simple}. We use automatically generated testbenches to evaluate the Verilog code produced by the model and collect preference pairs based on the test results. Through DPO, the model learns to generate code that is more likely to pass verification, improving its reliability and practical utility in hardware design tasks.

\subsection{Supervised Fine-Tuning with Realistic Specification}
\subsubsection{Specification Generation}
In hardware design workflows, engineers work with detailed specifications before implementing the Verilog code. Our specification format was developed with the input of experienced hardware engineers and includes comprehensive elements absent in typical academic datasets: detailed port descriptions, internal working principles, timing behaviors, and signal relationships. We use the following \raisebox{-0.1em}{\prompt{Generate Specification}} propmpt as shown in Section \ref{sec:prompt} to generate a realistic design specification $\x_i$ given the Verilog code $\y_i$. We denote the dataset for SFT as $\dsft = \{ (\x_i, \y_i) \}_{i=1}^N$.

\subsubsection{Training Objective}
We perform instruction tuning on foundation models $\pi_\phi$ with parameters $\phi$, to enhance the model's ability to respond to specifications. This process involves training on data sets $\dsft = \{ (\x_i, \y_i) \}_{i=1}^N$, consisting of the specification $\x_i$ paired with their corresponding verilog code $\y_i$. 
Specifically, maximum likelihood estimate (MLE) loss is used to find the best model parameters:
\begin{equation}
  \mathcal{L}_{\text{SFT}} = - \mathbb{E} \left[ \sum_{t=1}^{T} \log \pi_\phi \left( \y_{i, t} \mid \x_i, \y_{i, <t} \right) \right]  
\end{equation}
This objective minimizes the negative logarithmic likelihood of predicting each token $y_t$ given the specification $x$ and previous tokens $y_{<t}$, which is equivalent to minimizing the perplexity of the model on the training data.

\subsection{Reinforcement Learning with Testbench Feedback}
\subsubsection{Verification-Driven Testbench Generation}
\input{figs/tb_pipeline_cap}
The core principles for designing the testbench generation pipeline are decomposition and feedback. Decomposition means breaking down complex problems into smaller and more manageable subtasks, which prevents LLMs from being overwhelmed by complex problems \cite{lesat2most, decompose, drattack, haven}. We divide the testbench generation pipeline into four subtasks: \textbf{Analyze, Draft, Improve} and \textbf{Rectify}.
The goal of \textbf{Analyze} is to extract key functional requirements from the design specification and to determine the main test cases. \textbf{Draft} aims to generate an initial testbench that connects to the Verilog code $ \y_i $, implements the selected test cases, and records the verification results. In \textbf{Improve}, we enhance the testbench by analyzing coverage reports and modifying test cases to increase code coverage. The purpose of \textbf{Rectify} is to refine the verification conditions in the testbench by comparing their results with the actual behavior of the original code $ \y_i $, repeating this process if necessary to ensure correctness.

Feedback incorporates real environmental responses and external tool verification to reduce hallucinations \cite{hallucination, tool, react} and introduce verification insights.
For the last three subtasks, \textbf{Draft}, \textbf{Improve}, and \textbf{Rectify}, we use VCS, a simulation and debugging tool developed by Synopsys to verify digital designs described in Verilog, to provide different types of feedback. 

As shown in Figure \ref{fig:tb_pipeline}, the component {\prompt{xxx}} indicates calling the commercial LLM with the prompt, while component \vcs{xxx} represents the use of vcs tools to obtain feedback. Each feedback produces three possible statuses: \raisebox{-0.3em}{\includegraphics[height=1.2em]{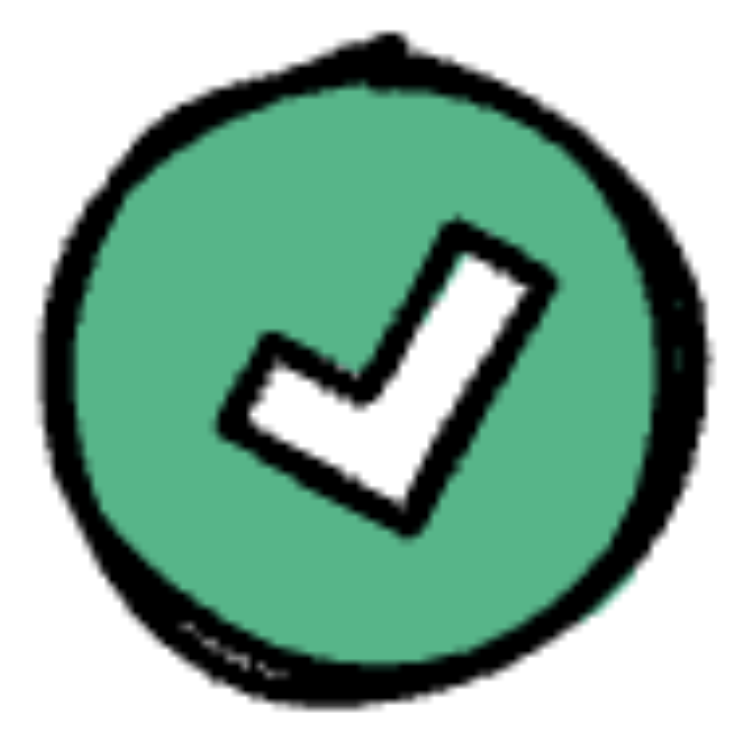}} indicates satisfying the requirement of this subtask and proceeding to the next subtask or \custombox{Finish} the pipeline; \raisebox{-0.3em}{\includegraphics[height=1.2em]{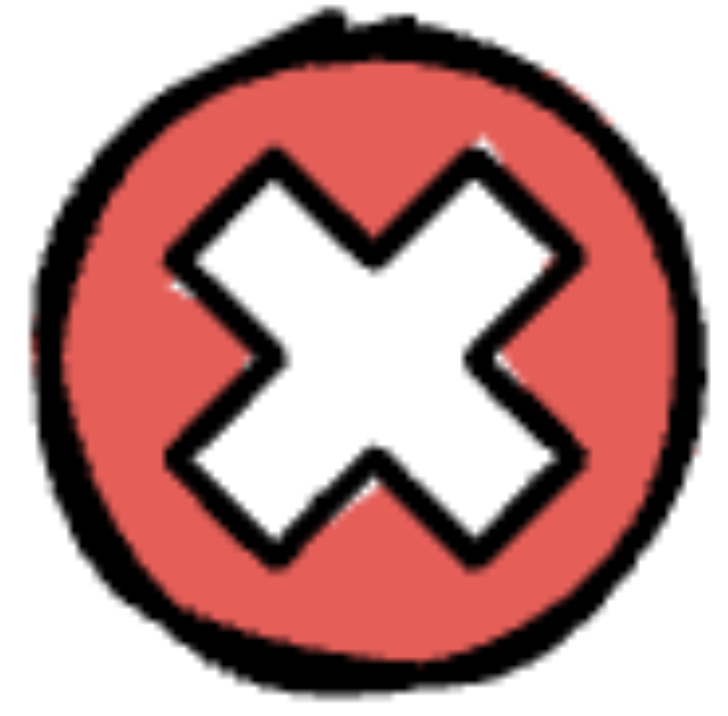}} indicates failure, requiring commercial LLMs to solve the problem; \raisebox{-0.3em}{\includegraphics[height=1.2em]{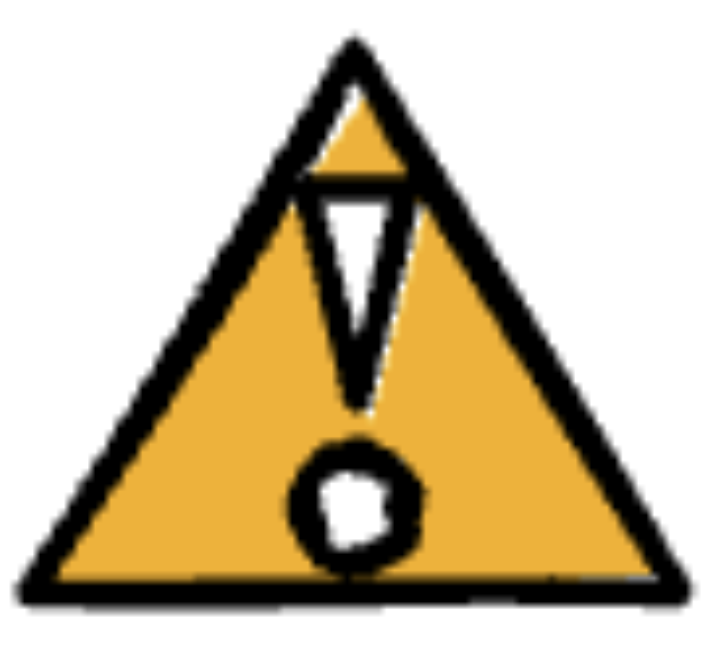}} indicates reaching the maximum attempts to solve the problem, at which point the pipeline will \custombox{Terminate} and discard this data instance for testbench generation. The components connected by concrete arrow is the critical path of this pipeline.

\circledblack{1} \textbf{Analyze}. The initial step analyzes the design specification and determines what should be tested. We first use the prompt \prompt{Generate Function Points} to guide the commercial LLM to identify important function points, which are key operational aspects of the module in various application scenarios. We provide only the specification, making the LLM focus on high-level understanding rather than on implementation details \cite{verilogcoder}. Subsequently, we use \prompt{Generate Test Cases} to identify important test cases, which are structured validation procedures with specific objectives, setup steps, and coverage dimensions. We limit the output to the top 3 function points and the top 5 test cases because unlimited generation results in repetitive content that affects subsequent stages. This limitation does not compromise the completeness of the testbench, since it suffices for our dataset, and additional testcases will be added to guarantee the coverage of the testbench in the \textbf{Improve} subtask.

\circledblack{2} \textbf{Draft}. We generate a testbench draft that establishes connections with the Verilog code $\y_i$ and configures stimuli for all test cases. Furthermore, as highlighted in {\setlength{\fboxrule}{1.5pt}\setlength{\fboxsep}{1pt}\color{red}\fbox{\color{black}red box}} within \prompt{Draft Testbench}, the testbench must record the verification results for each test case. As language models, LLMs cannot execute code to determine conditions and often suffer from hallucination. We will use these logged verification results in the later \textbf{Rectify} subtask. 
\input{prompts/draft_red}

We then use \vcs{Compile} to verify the successful compilation of the draft. If compilation fails, the error log is captured and returned along with the draft to \prompt{Draft Testbench}. The LLM then attempts to regenerate the draft. The pipeline \custombox{Terminate} after the maximum unsuccessful compilation attempts.

\circledblack{3} \textbf{Improve}. After generating the draft testbench, we employ \vcs{Coverage Report} to measure the line coverage of the Verilog code $\y_i$ when tested with our draft testbench.
\input{figs/report}

Figure \ref{fig:report} illustrates an example of such a report. When total line coverage exceeds the threshold, the testbench advances to the subsequent subtask. Otherwise, we append the coverage report to the prompt \raisebox{-0.1em}{\prompt{Improve Testbench}} to enhance the testbench's coverage percentage. Additionally, the pipeline \custombox{Terminate} if the improvement attempts reach the maximum limit.

\circledblack{4} \textbf{Rectify}. The final step involves rectifying the conditions in the testbench by running \vcs{Verify} on the Verilog code $\y_i$ under the testbench. In this stage, we treat the original Verilog code as the reference code to verify the testbench. 
Figure \ref{fig:simulation} shows how the verification logs reveal discrepancies between the conditions generated by LLM (first row) and the actual output of the code. When the verification log contains {\ttfamily{"Test completed with xx failure"}} as illustrated in Figure \ref{fig:simulation}, we append this log to the prompy \raisebox{-0.1em}{\prompt{Rectify Testbench}} prompt and resubmit the generation to \vcs{Verify}. If this rectification loop exceeds three iterations, the pipeline will be \custombox{Terminate}. In contrast, when the log indicates {\ttfamily{"Your Design Passed"}}, confirming that all testcase conditions in the generated testbench align with the Verilog code $\y_i$, we \custombox{Finish} the pipeline and consider testbench generation successful.
\input{figs/simulation}

Figure \ref{fig:tb} shows an example of generated testbench.
\input{figs/tb}

\subsubsection{Preference Pairs Collection}
\label{sec:collection}
We employ the fine-tuned model \oursft{} to generate two generated codes $\hat{\y}_i^1, \hat{\y}_i^2$ given the design specification $\x_i$. Subsequently, we use the generated testbench $\boldsymbol{v}_i$ to designate one generated code as preferred and the other as less preferred. When testing the generated code under the testbench, three possible statuses emerge: the code fails compilation, the code partially passes the testbench (including passing 0 cases), or the code fully passes the testbench. We discard pairs where either code fails to compile, as low-quality data can negatively impact learning performance, as demonstrated in previous research \cite{flow, plum}. This decision is further supported by our ablation study in Section \ref{sec:ablation}. The code that passes more testcases is considered preferred and is denoted $\yplus$.
We denote the final preference pairs dataset as $\drl=\{ (\x_i, \y_i^+, \y_i^-) \}_{i=1}^N$.

\subsubsection{Training Objective}

After collecting preference pairs $\drl
$, we use direct preference optimization (DPO) \cite{dpo}, a widely used reinforcement learning method, to further train our fine-tuned models \oursft{}. DPO works by directly optimizing the policy based on preference pairs without requiring an explicit reward model. The objective function of DPO is defined as
\begin{equation}
\mathcal{L}_{\rm RL} = -\mathbb{E}\left[\log\sigma\left(
\beta\log\frac{\pi_\theta(\yplus_i|\x_i)}{\pi_{\rho}(\yplus_i|\x_i)}
-\beta\log\frac{\pi_\theta(\yminus_i|\x_i)}{\pi_{\rho}(\yminus_i|\x_i)}
\right)\right]
\end{equation}

The loss encourages the model to increase the probability of preferred generated code ($\y^+$) while decreasing it for less preferred generated code ($\y^-$). Here, $\sigma$ is the sigmoid function, and $\beta$ is a temperature parameter controlling the optimization strength. $\pi_\theta$ represents the policy model that is being optimized during DPO training, while $\pi_\rho$ denotes the fixed reference model initialized from the fine-tuned model. This formulation enables learning from preference pairs while maintaining proximity to the reference model. Consequently, this conservative update strategy preserves the model's ability to respond to specifications while improving its Verilog code generation to satisfy functional correctness requirements in testbenches.

\section{Experimental Settings}

\subsection{Foundation Models}
Following previous works\cite{betterv,autovcoder,codev,haven}, we use CodeLlama-7B-Instruct \cite{codellama}, Deepseek-Coder-6.7B-Instruct \cite{deepseekcoder_v1.5}, and CodeQwen1.5-7B-Chat \cite{codeqwen} as foundation models. To evaluate the effectiveness of our method on models with MoE structures, we apply it to Mistral-7B-Instruct-v0.2 \cite{mistral}. Furthermore, we train our model using Qwen2.5-Coder-7B-Instruct and Qwen2.5-Coder-14B-Instruct, which belong to the Qwen2.5-Coder family, the most advanced code LLMs currently available.

\subsection{Datasets}
\subsubsection{Supervised Fine-tuning}
The verilog code is sourced from PyraNet \cite{pyranet}, a multilayered hierarchical dataset. In this dataset, data are synthesized using commercial LLMs and organized into four complexity tiers: Basic, Intermediate, Advanced, and Expert. In addition, the generated Verilog codes are self-evaluated by commercial LLMs, categorizing the data into six quality levels. The total dataset contains 692,238 entries. We further remove data with empty descriptions and retain only the top two quality levels, with 86,672 entries left.

\subsubsection{Reinforcement Learning}
\input{tables/num_pairs}
We extracted data from the SFT dataset $\dsft$ where the code exceeds 50 lines, as after SFT, the model demonstrates its ability to generate simple Verilog modules. This filtering process yielded 8,291 data instances. We selected GPT-4o as our commercial LLM. Using our pipeline, we generated 6,704 valid testbenches. We then used fine-tuned models \oursft{} with the design specification $\x$ with valid testbenches and used the testbench to collect preference pairs. Table \ref{tab:num_pairs} shows the number of preference pairs we collected for different \oursft{} variants. 

\subsection{Model Training}
The experiments were carried out using 8 Nvidia A100-80GB GPUs. The training process involved two stages: fine-tuning and alignment, both utilizing the AdamW optimizer. During fine-tuning, we employed a learning rate of $1\text{e-}5$ with a cosine learning rate scheduling and a warm-up ratio of 0.1 over 3 epochs. Subsequently, for alignment, we adjusted the learning rate to $5\text{e-}6$ while maintaining the same cosine learning rate scheduling and warm-up ratio of 0.1 through 10 epochs. The temperature parameter $\beta$ was set to 0.1, consistent with the default setting in the DPO paper\cite{dpo}.

\subsection{Model Inference}
During inference, only the design specification $x$ is available for the model. For our experiments, we used vLLM with specific configurations for the inference engine. The engine operates with the bf16 type and utilizes tensor parallelism across four devices, while maintaining a maximum token limit of 4096. We configure the sampling parameters with top\_p at 0.95 and top\_k at 50. Following the methodology established in previous work \cite{rtlcoder,autovcoder,haven}, we reported optimal results in three temperature: 0.2, 0.5, and 0.8.
\input{tables/results}

\subsection{Benchmarks}
We evaluated our models on the two most commonly used benchmarks in Verilog generation: RTLLM v1.1 \cite{rtllm} and VerilogEval \cite{verilogeval}. Both benchmarks require LLMs to generate RTL designs from natural language descriptions. RTLLM v1.1 comprises 29 RTL designs, 11 designs are arithmetic based, and 18 designs are logic based. VerilogEval consists of two components: VerilogEval-machine with 143 Verilog generation tasks created by GPT-based models, and VerilogEval-human with 156 manually crafted problems. Furthermore, we conduct additional evaluations on RTLLM v2 \cite{rtllm2}, which extends RTLLM v1.1 to 50 designs in four categories: Arithmetic, Memory, Control, and Miscellaneous. We also evaluate VerilogEval v2\cite{verilogeval2}, which builds upon VerilogEval-Human by implementing a chatbot-style format.

\subsection{Metrics}
We evaluate the models using the widely-adopted \pass$k$ metric for code generation, which is the percentage of problems solved by using $k$ generated programs per problem\cite{passnk}: 
\begin{equation}
    pass@k := \mathbb{E}_{i} \left[ 1 - \frac{\binom{n-c_i}{k}}{\binom{n}{k}} \right]
\end{equation}
where $n$ is the total number of trials for each specification and
$c_i$ is the number of correct code generations for task $i$. We set
$n=20$ in this experiment for comparison with baselines.
When any code within the $k$ trials successfully passes the test, we consider the task addressed. The \pass${k}$  metric therefore represents the estimated percentage of design tasks that can be successfully completed.
We measure the syntax and functional metrics \pass${1}$ and \pass${5}$, where \textbf{syntax} means that the code is compiling successfully, and \textbf{function} means that the code passes the testbench.

\section{Results and Analysis}
\subsection{Main Results}
Table \ref{tab:results} presents a comparison of our models with baselines, including general LLMs, code LLMs, and Verilog generation LLMs. In this study, \oursft{} represents the model after SFT, while \oursrl{} refers to the model further optimized through DPO applied to \oursft{}.

The evaluation is conducted on three benchmarks: VerilogEval-Machine, VerilogEval-Human and RTLLM v1.1. The baseline results are mainly sourced from \cite{haven}. To address the significant variance observed in the \pass$5$ metric introduced in \cite{rtlcoder}, we re-evaluated the open-source baselines and report unbiased \pass$k$. DeepSeek-v3 is excluded from the comparison due to its large size (671B), which makes a fair comparison infeasible.
\input{tables/verilogeval}

Our model achieves strong performance across all benchmarks, surpassing the baselines on VerilogEval-Human and RTLLM v1.1. The use of DPO improves performance across all variants $\text{\ours}$, regardless of their structures, model families, or advances in code LLMs. Although our model performs less effectively on the VerilogEval-Machine benchmark due to its focus on high-level human-like specifications during DPO training, $\text{\ours}_{\rm RL}$, based on Qwen2.5-Coder-14B, achieves improvements of $3\%$, $8.6\%$, and $7.8\%$ in \pass$1$ metrics on VerilogEval-Machine, VerilogEval-Human, and RTLLM v1.1, respectively, surpassing prior baselines.

We further evaluate our models, general LLMs, and open-source Verilog generation LLMs on two advanced benchmarks: RTLLM v2 and VerilogEval v2. Additionally, we assess the latest general LLM, DeepSeek-v3. Our models demonstrate superior performance on these benchmarks, as their prompts align more closely with the expectations of HDL engineers. While DeepSeek-v3, with 671B parameters, achieves exceptional results, $\text{\ours}_{\rm RL} $, based on Qwen2.5-Coder-14B with 14B parameters, delivers competitive results. Specifically, our model achieves improvements of $4.0\%$ and $7.3\%$ in the \pass$1$ metrics on VerilogEval v2 and RTLLM v2, respectively, compared to previous state-of-the-art results.

\input{tables/rtllm2}

\subsection{Generalization}
As shown in Tables \ref{tab:results}, \ref{tab:rtllm2}, and \ref{tab:verilogeval}, our method is generalized across different model structures, families, and advancements. 
\input{tables/verilogeval_general}

To further validate generalization of our method, we evaluated open-source Verilog generation LLMs, including RTLCoder \cite{rtlcoder}, CodeV \cite{codev}, \cite{haven}, and OriGen \cite{origen}, which were fine-tuned on their respective datasets. Using the same collection pipeline of preference pairs $\drl$ and the same settings as for our fine-tuned models, we applied DPO to these models while aligning the data format with their fine-tuning datasets. For simplicity, we tested only the DeepSeek-Coder variant without loss of generality.
\input{tables/rtllm2_general}

Tables \ref{tab:verilogeval_general} and \ref{tab:rtllm2_general} show that RL improves all Verilog generation LLMs on VerilogEval v2 and RTLLM v2. For example, RTLCoder achieves $2.4\%$ and $2.7\%$ improvements in \pass$1$ and \pass$5$ on VerilogEval v2, while OriGen shows $1.4\%$, $2.3\%$ and $2.1\%$ improvements in \pass$1$, \pass$5$ and \pass$10$ on RTLLM v2.

\subsection{Ablation Study}
\subsubsection{Practical Specification}

We perform the fine-tuning on the original design specification. Figure~\ref{fig:simple_spec} shows an example. As shown in the "Simple" row in Table~\ref{tab:ablation}, training with the simple design specification achieves poor results compared to \oursft{}.

\subsubsection{SFT with Verification Insight}
Here we use SFT with verification data. We keep the preferred code that passes the testbench and pair it with the design specification to continue fine-tuning \oursft{}. This process, known as reject sampling \cite{reject_sampling}, means that only samples that meet certain criteria, such as passing some tests on the testbench, are selected for further fine-tuning. However, as shown in the "Verification Insights" in Table~\ref{tab:ablation}, SFT with verification insights does not show a consistent improvement.
\input{figs/simple_spec}

\subsubsection{Preference Pair Construction}
\label{sec:ablation}
In this section, we demonstrate the importance of using the testbench to generate preference pairs. We conducted an ablation study by constructing preference pairs using bilingual evaluation understudy (BLEU), abstract syntax tree (AST), and data flow graph (DFG).

The BLEU \cite{bleu} score evaluates the similarity between the generated code and the verilog code by comparing their overlapped $n$-grams. We use 4-grams with uniform weights, and a brevity penalty is applied to prevent overly short candidate sentences from achieving high scores. For AST, we follow the approach in VeriSeek\cite{veriseek}, which uses AST similarity as a reward to guide PPO. 
For DFG, we utilize Pyverilog \cite{pyverilog} to parse the graph $D$ and use Jaccard similarity:
\begin{equation}
\mathcal{J}(D(\yhat), D(\y)) = \frac{|D(\yhat) \cap D(\y)|}{|D(\yhat) \cup D(\y)|}
\end{equation}
where $\yhat$ represents the generated code; $\y$ denotes the original code.

All preference pairs are constructed only if both generations can be compiled successfully. The generated code with a higher similarity to the verilog code is treated as the preferred code. Table \ref{tab:ablation} shows that while methods such as BLEU improve some metrics (e.g., \pass$1$ on RTLLM v2), they lack consistent performance improvements and often cause decreases in other metrics.

Furthermore, we examine whether treating generated code with syntax errors as $\yminus$ and syntax-valid generated codes as $\yplus$ can improve model performance. Unfortunately, Table \ref{tab:ablation} shows that training with syntax errors code reduces performance, supporting the choice in the above section.\ref{sec:collection}.

Finally, the performance differences are not caused by the size of the dataset $\drl$ used for DPO. The testbench-based approach generates the fewest preference pairs, yet achieves the best results.

\subsubsection{Reinforcement Algorithms}

A previous work VeriSeek \cite{veriseek} uses a parallel structure-aware AST-based reward to guide PPO training. In this ablation study, we investigated the use of DPO pair construction methods to guide PPO. Specifically, we directly applied the similarity of BLEU, AST, and DFG, as their similarity values are in $[0,1]$ by default. For the testbench, we set the reward as the proportion of test cases passed. Consequently, all rewards fall within $[0,1]$, which satisfies the requirements of Algorithm.1 in \cite{veriseek}.

However, the results in Table \ref{tab:ablation} show that this approach reduces performance compared to SFT. This is likely due to reward hacking\cite{rewardhacking}, where the model is over-fitted to maximize the proportion of passed test cases without improving model quality. In contrast, DPO avoids this issue by using pairwise comparisons from testbench outcomes, aligning outputs with hardware design task without relying on an explicit reward function.
\input{tables/ablation}

\subsection{PPL versus Functional Correctness}
\label{sec:ppl}
Current Verilog generation LLMs are almost all trained with SFT, which minimizes perplexity (PPL). Although a low PPL on an evaluation set indicates that the LLM has learned the data distribution well, it does not guarantee generating code that is functionally correct.

Figure~\ref{fig:misalign} shows that PPL does not correlate with functional correctness. For code pairs with the same design specification, the x-axis shows the PPL of the code passing more testcases, and the y-axis shows the other. Points above the diagonal mean better-performing code has lower PPL; below means higher PPL. Only 52.3\% of pairs show that passing more testcases corresponds to lower PPL.
\input{figs/ppl}

\section{Conclusion}
In this work, we address the misalignment between Verilog generation LLMs and actual hardware design requirements by incorporating verification feedback into the training process. Our method combines an automatic testbench generation pipeline with DPO, enabling the model to learn from functional correctness signals rather than relying on explicit reward functions. Experimental results across multiple benchmarks confirm that our approach consistently improves the functional correctness of generated Verilog code and outperforms existing baselines. By integrating verification feedback and reinforcement learning, we provide a practical solution to generate Verilog code that is more likely to be functionally correct.

\section{Detailed Prompts}
\label{sec:prompt}
\input{prompts/spec}
\input{prompts/improve}
\input{prompts/funcpoint}
\input{prompts/testcase}
\input{prompts/draft}
\input{prompts/rectify}
\clearpage

\bibliographystyle{abbrv}
\bibliography{ref}

\end{document}

%% file: figs/simple_cap.tex
\begin{figure}[t]
    \centering
    \includegraphics[width=\linewidth]{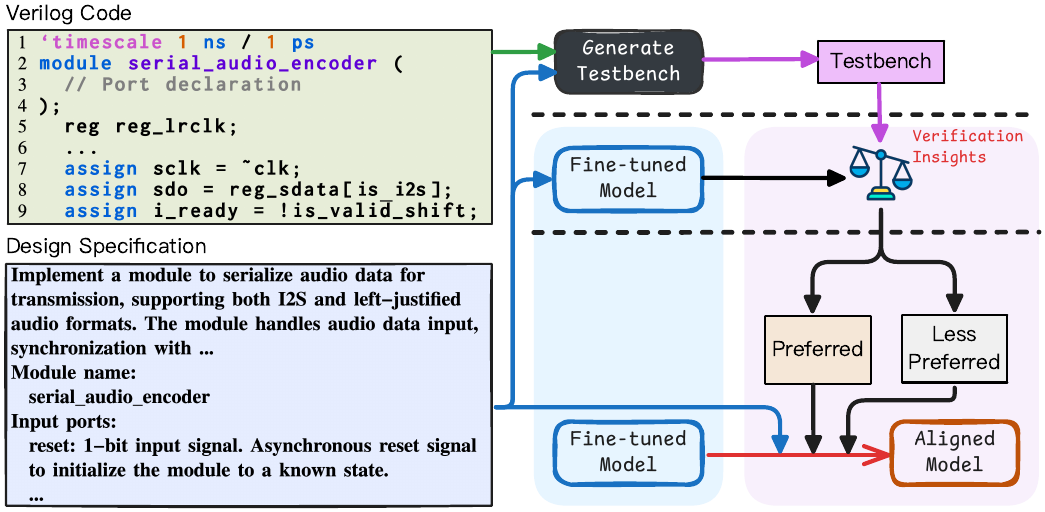}
    \caption{Overview of our work. We first use paired design specification and Verilog code to automatically generate testbenches. Then we prompt the fine-tuned model and test the generated code using the testbenches to collect verification insights. The code that passes more testcases is considered as preferred, and the other as less preferred. Finally, the design specification with the preference pairs is used for reinforcement learning.
    }
    \label{fig:simple}
\end{figure}

%% file: figs/tb_pipeline_cap.tex
\begin{figure}[t]
    \centering
    \includegraphics[width=\linewidth]{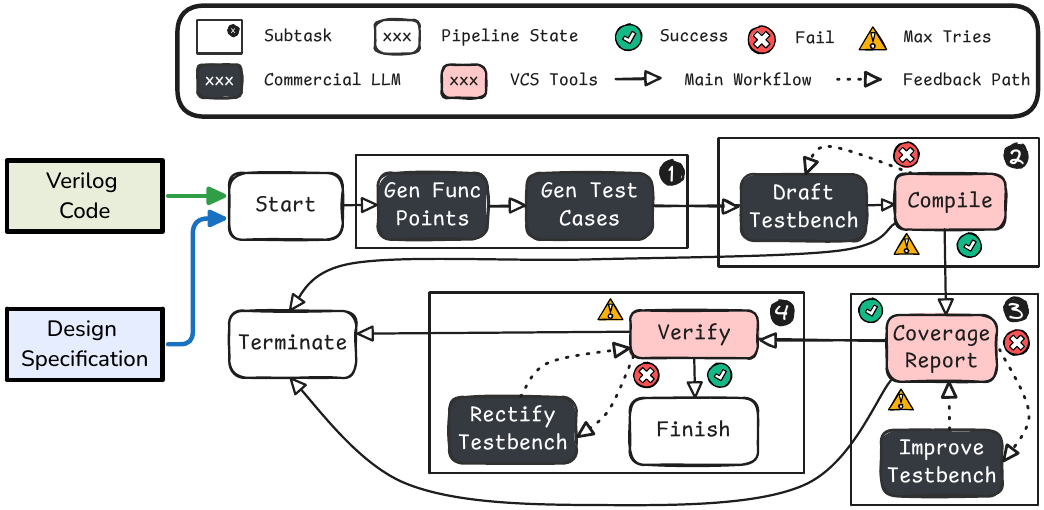}
    \caption{Automatic testbench generation pipeline.
    }
    \label{fig:tb_pipeline}
\end{figure}

%% file: prompts/draft_red.tex
\begin{tcolorbox}[title=Draft Testbench, left=0mm, right=0mm, top=0mm, bottom=0mm,  colback=white, fonttitle=\footnotesize]
\begin{lstlisting}[style=verilog]
  ...
  // Must set stimulus and report all testcases with simulation results
  $display("===========TestCases===========")
  // Test Case 1:
  $display("Test Case 1. Expected x1: [VALUE], ...")
  (*@{\setlength{\fboxrule}{1.5pt}\setlength{\fboxsep}{1pt}\color{red}\fbox{\color{black}\$display("Test Case 1. Actual x1: \%d, ...", x1, ...)}}@*)
  ...
  $display("===========End===========")
  ...
\end{lstlisting}
\end{tcolorbox}

%% file: figs/report.tex
\begin{figure}[h]
\centering
\begin{tcolorbox}[left=0mm, right=0mm, top=0mm, bottom=0mm,  colback=white, colbacktitle=blue, colframe=black, fonttitle=\bfseries, width=0.9\linewidth]
\begin{lstlisting}[style=plain]
Since this is the module's only instance, the coverage report is the same as for the module.
Line Coverage for Module : serial_audio_encoder
Line No.	Total	Covered	Percent
TOTAL		31	26	(*@{\setlength{\fboxrule}{1.5pt}\setlength{\fboxsep}{1pt}\color{red}\fbox{\color{black}83.87}}@*)
CONT_ASSIGN	25	1	1	100.00
...
\end{lstlisting}

\begin{lstlisting}[style=report]
    ...
    always @(posedge clk or posedge reset) begin
1/1   if (reset) begin
        ...
      end else begin
1/1     if (is_valid_shift) begin
1/1       shift_count <= shift_count - 1'b1;
1/1       is_valid_shift <= shift_count != 0;
1/1       shift <= shift << 1;
          ...
        end else begin
0/1 ==>    reg_lrclk <= 1'b1;
0/1 ==>    is_next_left <= 1'b1;
0/1 ==>    is_valid_shift <= 1'b0;
0/1 ==>    reg_sdata <= 2'b00;
0/1 ==>    is_underrun <= 1'b1;
     ...
\end{lstlisting}
\end{tcolorbox}
\caption{Line coverage report example. {\setlength{\fboxrule}{1.5pt}\setlength{\fboxsep}{1pt}\color{red}\fbox{\color{black}Red box}} indicates the total line coverage percentage. {\ttfamily{1/1}} before the line means it is covered by the testbench and {\ttfamily{\color{red}{0/1}}} means not.}
\label{fig:report}
\end{figure}

%% file: figs/simulation.tex
\begin{figure}[htbp]
\centering
\begin{tcolorbox}[left=1mm, right=1mm, top=0.5mm, bottom=0.5mm,  colback=white, colbacktitle=blue, colframe=black, fonttitle=\bfseries, width=0.9\linewidth]
\begin{lstlisting}[style=plain]
Test Case 1. Expected i_ready: 2'b01, is_underrun: 3'b100
Test Case 1. Actual i_ready: 0, is_underrun: 0
...
===========End===========
Test with 5 failures
\end{lstlisting}
\end{tcolorbox}
\caption{Simulation output example.}
\label{fig:simulation}
\end{figure}

%% file: figs/tb.tex
\begin{figure}[h]
\centering
\begin{tcolorbox}[left=0mm, right=0mm, top=0mm, bottom=0mm,  colback=white, colbacktitle=blue, colframe=black, fonttitle=\bfseries, width=0.95\linewidth]
\begin{lstlisting}[style=verilog]
`timescale 1ns / 1ps
module testbench;
// Parameters
localparam width = 16;
// Inputs
reg reset;
...
// Instantiate the Unit Under Test (UUT)
serial_audio_encoder #(width) uut (
  ...
);
// Clock generation
always #5 clk = ~clk;
initial begin
  ...
  // Test Case 1: Reset during operation
  reset = 1;
  #10;
  reset = 0;
  #10;
  $display("Test Case 1. Expected i_ready: 0, is_underrun: 0");
  $display("Test Case 1. Actual i_ready: %b, is_underrun: %b", i_ready, is_underrun);
  if (i_ready !== 0 || is_underrun !== 0) error_count = error_count + 1;
  ...
  // Must finish simulation
  $finish;
end
endmodule

\end{lstlisting}
\end{tcolorbox}
\caption{An example of generated testbench.}
\label{fig:tb}
\end{figure}

%% file: tables/num_pairs.tex
\begin{wraptable}{r}{0.45\linewidth}
\footnotesize
\caption{Number of pairs for \oursft{} variants.}
\setlength{\tabcolsep}{5pt}
\label{tab:num_pairs}
\centering
\begin{tabular}{@{}cc@{}}
\toprule
\textbf{\oursft{}} & \textbf{\#Pairs} \\ \midrule
\textbf{Mistral-v0.2} & 1782 \\
\textbf{CodeLlama} & 1991 \\
\textbf{DeepSeek-Coder} & 2170 \\
\textbf{CodeQwen} & 2223 \\
\textbf{Qwen2.5-Coder} & 1796  \\
\textbf{Qwen2.5-Coder-14B} & 2267 \\ \bottomrule
\end{tabular}
\end{wraptable}

%% file: tables/results.tex
\begin{table*}[t]
\centering
\caption{Comparison of Model Performance On VerilogEval-Machine, VerilogEval-Human and RTLLM v1.1.}
\label{tab:results}
\begin{threeparttable}
\setlength{\tabcolsep}{2.5pt}
\footnotesize
\begin{tabular}{cc c c cccc cccc}
\toprule
& & & & \multicolumn{2}{c}{\textbf{VerilogEval-Machine}} & \multicolumn{2}{c}{\textbf{VerilogEval-Human}} & \multicolumn{4}{c}{\textbf{RTLLM v1.1}} \\
\cmidrule(lr){5-8} \cmidrule(lr){9-12}
& & & & \multicolumn{2}{c}{\textbf{Function}} & \multicolumn{2}{c}{\textbf{Function}} & \multicolumn{2}{c}{\textbf{Syntax}} & \multicolumn{2}{c}{\textbf{Function}} \\
\cmidrule(lr){5-6} \cmidrule(lr){7-8} \cmidrule(lr){9-10} \cmidrule(lr){11-12}
\multirow{-3}{*}{\textbf{Type}} &
\multirow{-3}{*}{\textbf{Model}} &
\multirow{-3}{*}{\begin{tabular}[c]{@{}c@{}}\textbf{Open} \\\textbf{Source}\end{tabular}} &
\multirow{-3}{*}{\textbf{\#Params}} &
\pass${1}$ &
\pass${5}$ &
\pass${1}$ &
\pass${5}$ &
\pass${1}$ &
\pass${5}$ &
\pass${1}$ &
\pass${5}$
\\ \midrule

\multirow{3}{*}{\textbf{General LLMs}}
& GPT-4o\cite{gpt4o} & \ding{55} & N/A & 65.9 & 71.4 & \cellcolor{bronze}57.1 & 63.9 & 82.4 &86.2 & 47.9 & 58.0 \\
& Claude-3.5-Sonnet\cite{claude} & \ding{55} & N/A & 60.2 & 75.5 & 46.1 & 56.0 & 90.0 & \cellcolor{silver}99.9 & \cellcolor{silver}51.1 & 60.0 \\
& Llama3.1 \cite{llama3} & \ding{51} & 405B & 67.3 & 75.1 & 53.8 & 61.0 & 73.2 & 81.8 & 38.9 & 45.8 \\
\midrule

\multirow{6}{*}{\textbf{Coding LLMs}} 
& Mistral-v0.2 \cite{mistral} & \ding{51} & 7B & 36.9 & 48.8 & 21.1 & 27.1 & 5.3 & 14.2 & 2.2 & 8.0 \\
& CodeLlama \cite{codellama} & \ding{51} & 7B & 43.1 & 47.1 & 18.2 & 22.7 & 46.6 & 62.6 & 17.9 & 29.9 \\
& DeepSeek-Coder \cite{deepseekcoder_v1.5} & \ding{51} & 6.7B & 52.2 & 55.4 & 30.2 & 33.9 & 51.4 & 64.4 & 23.1 & 29.3 \\
& CodeQwen \cite{codeqwen} & \ding{51} & 7B & 46.5 & 54.9 & 22.5 & 26.1 & 45.8 & 65.8 & 24.1 & 34.0 \\
& Qwen2.5-Coder \cite{codeqwen2} & \ding{51} & 7B & 59.9 & 69.7 & 33.6 & 41.4 & 82.2 & 94.8 & 28.6 & 44.4 \\
& Qwen2.5-Coder-14B \cite{codeqwen2} & \ding{51} & 14B & 60.3 & 66.1 & 44.4 & 52.2 & 86.0 & 92.2 & 39.5 & 50.0 \\
\midrule

\multirow{-1}{*}{\textbf{VeriGen}\cite{verigen}} & CodeGen & \ding{51} & 16B & 35.6 & 51.7 & 23.0 & 39.7 & 86.2 & 90.2 & 24.1 & 34.6 \\ \midrule

& Mistral & \ding{51} & 7B & 62.5 & 72.2 & 36.7 & 45.5 & 64.6 & 73.7 & 24.5 & 37.3 \\
\multirow{-2}{*}{\textbf{RTLCoder}\cite{verigen}} & DeepSeek-Coder & \ding{51} & 6.7B & 61.2 & 76.5 & 41.6 & 50.1 & 73.4 & 83.9 & 35.8 & 40.3 \\ \midrule

& CodeLlama & \ding{55} & 7B & 64.2 & 75.4 & 40.9 & 50.0 & N/A & N/A & N/A & N/A \\
& DeepSeek-Coder & \ding{55} & 6.7B & 67.8 & 79.1 & 45.9 & 53.3 & N/A & N/A & N/A & N/A \\
\multirow{-3}{*}{\textbf{BetterV}\cite{betterv}} & CodeQwen & \ding{55} & 7B & 68.1 & 79.4 & 46.1 & 53.7 & N/A & N/A & N/A & N/A \\ \midrule

& CodeLlama & \ding{55} & 7B & 63.7 & 72.9 & 44.5 & 52.8 & N/A & 100.0 & N/A & 48.3 \\
& DeepSeek-Coder & \ding{55} & 6.7B & 69.0 & 79.3 & 46.9 & 53.7 & N/A & \cellcolor{gold}100.0 & N/A & 51.7 \\
\multirow{-3}{*}{\textbf{AutoVCoder}\cite{autovcoder}} & CodeQwen & \ding{55} & 7B & 68.7 & 79.9 & 48.5 & 55.9 & N/A & 93.1 & N/A & 51.7\\ \midrule

& CodeLlama & \ding{51} & 7B & 78.1 & 86.0 & 45.2 & 59.5 & 79.0 & 89.2 & 39.4 & 50.3 \\
& DeepSeek-Coder & \ding{51} & 6.7B & 77.9 & \cellcolor{gold}{88.6} & 52.7 & 62.5 & 78.3 & 87.4 & 42.4 & 51.5 \\
\multirow{-3}{*}{\textbf{CodeV}\cite{codev}} & CodeQwen & \ding{51} & 7B & 77.6 & \cellcolor{silver}88.2 & 53.2 & \cellcolor{bronze}65.1 & 78.8 & 89.5 & 36.6 & 53.3 \\ \midrule

& CodeLlama & \ding{51} & 7B & 74.7 & 80.0 & 47.5 & 54.6 & 45.5 & 95.4 & 42.3 & 46.8 \\
& DeepSeek-Coder & \ding{51} & 6.7B & \cellcolor{bronze}78.8 & 84.5 & 46.6 & 56.6 & 88.9 & 92.8 & 45.4 & 55.3 \\
\multirow{-3}{*}{\textbf{HaVen}\cite{haven}} & CodeQwen & \ding{51} & 7B & 77.3 & 81.2 & 53.3 & 57.8 & 87.6 & 92.8 & 45.1 & 53.3 \\ \midrule

\multirow{-1}{*}{\textbf{OriGen}\cite{origen}} & DeepSeek-Coder & \ding{51} & 6.7B & 74.1 & 82.4 & 43.3 & 46.4 & 78.1 & 86.4 & 45.2 & 58.4 \\ \midrule

\multirow{6}{*}{\oursft{}} 
& Mistral-v0.2 & \ding{51} & 7B & 68.6 & 76.6 & 48.8 & 51.5 & 77.4 & 93.5 & 26.5 & 42.1 \\
& CodeLlama & \ding{51} & 7B & 70.1 & 82.5 & 44.7 & 55.6 & 82.6 & 93.2 & 33.0 & 50.4 \\
& DeepSeek-Coder & \ding{51} & 6.7B & 71.4 & 81.6 & 49.7 & 59.5 & 85.5 & 95.3 & 50.4 & 60.6 \\ 
& CodeQwen & \ding{51} & 7B & 76.1 & 84.7 & 43.3 & 58.0 & 87.6 & 96.3 & 38.7 & 58.4 \\
& Qwen2.5-Coder & \ding{51} & 7B & 72.1 & 84.8 & 46.7 & 61.4 & 87.9 & 96.4 & 50.0 & 62.1 \\
& Qwen2.5-Coder-14B & \ding{51} & 14B & \cellcolor{gold}{82.0} & \cellcolor{bronze}87.7 & \cellcolor{silver}61.1 & \cellcolor{silver}70.6 & \cellcolor{silver}92.1 & 95.9 & 47.6 & 63.1 \\
\midrule

\multirow{6}{*}{\oursrl{}}
& Mistral-v0.2 & \ding{51} & 7B & 68.8\plus{0.2} & 76.1\minus{0.5} & 50.2\plus{1.4} & 53.4\plus{1.9} & 76.0 & 88.7 & 30.5\plus{4.0} & 44.3\plus{2.2} \\
& CodeLlama & \ding{51} & 7B & 70.0\minus{0.1} & 81.9\minus{0.6} & 46.6\plus{1.9} & 55.5\minus{0.1} & 82.2 & 94.7 & 35.6\plus{2.6} & 53.0\plus{2.6} \\
& DeepSeek-Coder & \ding{51} & 6.7B & 71.2\minus{0.2} & 81.5\minus{0.1} & 52.8\plus{3.1} & 64.0\plus{4.5} & 87.4 & 96.1 & \cellcolor{bronze}50.7\plus{0.5} & \cellcolor{silver}64.7\plus{4.1} \\
& CodeQwen & \ding{51} & 7B & 75.0\minus{1.1} & 86.0\plus{1.3} & 44.6\plus{1.3} & 59.0\plus{1.0} & 87.2 & \cellcolor{bronze}96.5 & 39.0\plus{0.3} & 59.3\plus{0.9} \\
& Qwen2.5-Coder & \ding{51} & 7B & 72.7\plus{0.6} & 85.8\plus{1.0} & 49.7\plus{3.0} & 62.3\plus{0.9} & 90.1 & 93.8 & \cellcolor{gold}53.2\plus{3.2} & \cellcolor{gold}67.7\plus{5.6} \\
& Qwen2.5-Coder-14B & \ding{51} & 14B & \cellcolor{silver}81.8\minus{0.2} & 87.3\minus{0.4} & \cellcolor{gold}61.9\plus{0.8} & \cellcolor{gold}71.3\plus{0.6} & \cellcolor{bronze}90.5 & 95.2 & 50.2\plus{2.6} & \cellcolor{bronze}63.9\plus{0.8} \\

\bottomrule

\end{tabular}
\begin{tablenotes}\footnotesize
\item[+] The background colors \colorbox{gold}{Gold}, \colorbox{silver}{Silver} and \colorbox{bronze}{Bronze} denote the first, second, and third rankings, respectively.
\end{tablenotes}
\end{threeparttable}
\end{table*}

%% file: tables/verilogeval.tex
\begin{table}[t]
\centering
\caption{Comparison of Model Performance On VerilogEval v2.}
\label{tab:verilogeval}
\begin{threeparttable}
\setlength{\tabcolsep}{0.8pt}
\footnotesize
\begin{tabular}{cc cc cc}
\toprule
& & \multicolumn{4}{c}{\textbf{VerilogEval v2}} \\
\cmidrule(lr){3-6}
& & \multicolumn{2}{c}{\textbf{Syntax}} & \multicolumn{2}{c}{\textbf{Function}} \\
\cmidrule(lr){3-4} \cmidrule(lr){5-6}
\multirow{-3}{*}{\textbf{Type}} &
\multirow{-3}{*}{\textbf{Model}} &
\pass{$1$} & \pass{$5$} & \pass{$1$} & \pass{$5$} \\
\midrule

\multirow{4}{*}{\begin{tabular}[c]{@{}c@{}}\textbf{General}\\\textbf{LLMs}\end{tabular}} & GPT-4o & 90.7 & 95.2 & 56.5 & 65.1 \\
& Claude-3.5-Sonnet & 92.7 & 96.7 & 55.1 & 65.4 \\
& Llama3.1 & 91.2 & 93.5 & 50.2 & 57.1 \\
& Deepseek-v3-671B \cite{deepseekv3} & 96.4 & 99.6 & \cellcolor{gold}66.7 & \cellcolor{gold}74.0 \\
\midrule

\multirow{6}{*}{\begin{tabular}[c]{@{}c@{}}\textbf{Coding}\\\textbf{LLMs}\end{tabular}} & Qwen2.5-Coder & 82.9 & 94.5 & 34.4 & 43.8 \\
& Mistral & 16.2 & 34.0 & 6.6 & 9.7 \\
& CodeLlama & 37.1 & 56.1& 12.1 & 20.2 \\
& CodeQwen & 52.4 & 57.0 & 19.8 & 24.3 \\
& DeepSeek-Coder & 63.1 & 68.7 & 22.6 & 29.6 \\
& Qwen2.5-Coder-14B & 85.3 & 97.9 & 43.9 & 56.6\\
\midrule

& Mistral & 69.8 & 78.1 & 34.0 & 37.5 \\
\multirow{-2}{*}{\textbf{RTLCoder}} & DeepSeek-Coder & 87.0 & 94.6 & 40.9 & 47.9 \\
\midrule

& CodeLlama & 93.7 & 95.2 & 46.1 & 52.5 \\
& DeepSeek-Coder & 93.3 & 96.8 & 47.1 & 54.4 \\
\multirow{-3}{*}{\textbf{CodeV}} & CodeQwen & 95.4 & 95.9 & 48.4 & 56.1 \\
\midrule

& CodeLlama & 88.2 & 94.8 & 41.8 & 47.9  \\
& DeepSeek-Coder & \cellcolor{bronze}97.1 & 98.9 & 47.2 & 52.3 \\
\multirow{-3}{*}{\textbf{HaVen}} & CodeQwen & 97.0 & 98.5 & 47.6 & 51.9 \\
\midrule

\multirow{-1}{*}{\textbf{OriGen}} & DeepSeek-Coder & 77.5 & 87.8 & 51.3 & 56.3 \\
\midrule

\multirow{6}{*}{\oursft{}} 
& Mistral & 93.7 & 97.6 & 43.3 & 48.4 \\ 
& CodeLlama & 90.8 & 99.9 & 33.3 & 49.7 \\ 
& DeepSeek-Coder & \cellcolor{gold}98.5 & 99.9 & 50.1 & 57.6 \\
& CodeQwen & 95.6 & 99.9 & 45.8 & 60.7 \\ 
& Qwen2.5-Coder & 95.1 & \cellcolor{gold}100.0 & 47.9 & 64.3\\
& Qwen2.5-Coder-14B & 96.6 & 99.9 & \cellcolor{bronze}57.9 & \cellcolor{bronze}71.1\\
\midrule

\multirow{6}{*}{\oursrl{}} 
& Mistral & 92.0 & 96.5 & 44.4\plus{1.1} & 48.7\plus{0.3} \\ 
& CodeLlama & 91.3 & 99.8 & 35.3\plus{2.0} & 51.0\plus{1.3} \\ 
& DeepSeek-Coder & \cellcolor{silver}98.3 & \cellcolor{bronze}99.8 & 52.1\plus{2.0} & 58.5\plus{0.9} \\
& CodeQwen & 95.5 & 99.9 & 46.0\plus{0.2} & 60.8\plus{0.1} \\
& Qwen2.5-Coder & 95.6 & 99.9 & 49.3\plus{1.4} & 64.7\plus{0.4} \\
& Qwen2.5-Coder-14B & 96.8 & \cellcolor{silver}99.9 & \cellcolor{silver}58.6\plus{0.7} & \cellcolor{silver}72.1\plus{1.0}\\
\bottomrule

\end{tabular}
\begin{tablenotes}\footnotesize
\item[+] The background colors \colorbox{gold}{Gold}, \colorbox{silver}{Silver} and \colorbox{bronze}{Bronze} denote the first, second, and third rankings, respectively.
\end{tablenotes}
\end{threeparttable}
\end{table}

%% file: tables/rtllm2.tex
\begin{table}[htbp]
\centering
\caption{Comparison of Model Performance On RTLLM v2.}
\label{tab:rtllm2}
\begin{threeparttable}
\setlength{\tabcolsep}{0.8pt}
\footnotesize
\begin{tabular}{cc cc cc}
\toprule
& & \multicolumn{4}{c}{\textbf{RTLLM v2}} \\
\cmidrule(lr){3-6}
& & \multicolumn{2}{c}{\textbf{Syntax}} & \multicolumn{2}{c}{\textbf{Function}} \\
\cmidrule(lr){3-4} \cmidrule(lr){5-6}
\multirow{-3}{*}{\textbf{Type}} &
\multirow{-3}{*}{\textbf{Model}} &
\pass{$1$} & \pass{$5$} & \pass{$1$} & \pass{$5$} \\
\midrule

\multirow{4}{*}{\begin{tabular}[c]{@{}c@{}}\textbf{General} \\ \textbf{LLMs}\end{tabular}} & GPT-4o & 80.0 & 89.5 & 47.9 & 58.0 \\
& Claude-3.5-Sonnet & 83.8 & 95.4 & 44.0 & 57.6 \\
& Llama3.1 & 73.0 & 81.6 & 38.8 & 45.6 \\
& Deepseek-v3-671B \cite{deepseekv3} & 92.3 & 99.2 & \cellcolor{gold}56.8 & \cellcolor{gold}70.3 \\
\midrule

\multirow{6}{*}{\begin{tabular}[c]{@{}c@{}}\textbf{Code} \\ \textbf{LLMs}\end{tabular}} 
& Mistral-v0.2 & 11.3 & 28.1 & 2.5 & 6.8 \\ 
& CodeLlama & 49.5 & 76.7 & 21.2 & 31.9 \\
& CodeQwen & 47.1 & 66.4 & 25.8 & 29.0 \\
& DeepSeek-Coder & 72.7 & 88.1 & 26.5 & 36.3 \\
& Qwen2.5-Coder & 85.2 & 96.5 & 40.5 & 49.8 \\
& Qwen2.5-Coder-14B & 86.7 & 97.7 & 47.8 & 56.6 \\
\midrule

& Mistral & 75.6 & 81.1 & 37.0 & 39.9 \\
\multirow{-2}{*}{\textbf{RTLCoder}} & DeepSeek-Coder & 82.9 & 90.0 & 43.5 & 48.0 \\
\midrule

& CodeLlama & 76.8 & 91.3 & 47.4 & 53.3 \\
& DeepSeek-Coder & 75.5 & 90.7 & 45.5 & 55.7 \\
\multirow{-3}{*}{\textbf{CodeV}} & CodeQwen & 78.1 & 89.6 & 48.1 & 56.9 \\
\midrule

& CodeLlama & 51.7 & 92.4 & 44.9 & 48.2 \\
& DeepSeek-Coder & 89.0 & 94.2 & 50.6 & 60.5 \\
\multirow{-3}{*}{\textbf{HaVen}} & CodeQwen & 88.6 & 94.9 & 51.2 & 59.7 \\
\midrule

\multirow{-1}{*}{\textbf{OriGen}} & DeepSeek-Coder & 77.5 & 87.8 & 40.9 & 57.1 \\
\midrule

\multirow{6}{*}{\ours$_{\rm SFT}$} 
& Mistral-v0.2 & 83.3 & 96.5 & 32.5 & 51.9 \\ 
& CodeLlama & 86.7 & 98.2 & 38.0 & 56.6 \\ 
& DeepSeek-Coder & 88.6 & 96.9 & 48.1 & 59.9 \\ 
& CodeQwen & \cellcolor{silver}93.9 & \cellcolor{bronze}99.4 & 46.9 & 62.7 \\
& Qwen2.5-Coder & 92.3 & 99.8 & 49.8 & 64.5 \\
& Qwen2.5-Coder-14B & \cellcolor{gold}95.2 & 99.8 & \cellcolor{bronze}53.8 & 67.5 \\
\midrule

\multirow{6}{*}{\ours$_{\rm DPO}$} 
& Mistral-v0.2 & 79.0 & 95.2 & 35.2\plus{2.7} & 55.7\plus{3.8} \\ 
& CodeLlama & 88.0 & 99.2 & 39.2\plus{1.2} & 58.3\plus{1.7} \\
& DeepSeek-Coder & 90.6 & 97.6 & 50.4\plus{2.3} & 63.0\plus{3.1} \\
& CodeQwen & \cellcolor{bronze}92.7 & \cellcolor{gold}99.8 & 47.7\plus{0.8} & 62.7\plus{0.0} \\
& Qwen2.5-Coder & 90.1 & 99.6 & 52.4\plus{2.6} & \cellcolor{bronze}66.4\plus{1.9} \\
& Qwen2.5-Coder-14B & 85.2 & \cellcolor{silver}99.7 & \cellcolor{silver}55.2\plus{1.4} & \cellcolor{silver}67.5\plus{0.0} \\
\bottomrule

\end{tabular}
\begin{tablenotes}\footnotesize
\item[+] The background colors \colorbox{gold}{Gold}, \colorbox{silver}{Silver} and \colorbox{bronze}{Bronze} denote the first, second, and third rankings, respectively.
\end{tablenotes}
\end{threeparttable}
\end{table}

%% file: tables/verilogeval_general.tex
\begin{table}[htbp]
\centering
\caption{Aligned Verilog Generation LLMs Performance on VerilogEval v2}
\label{tab:verilogeval_general}
\begin{threeparttable}
\setlength{\tabcolsep}{2pt}
\footnotesize
\begin{tabular}{c c ccc ccc}
\toprule
& & \multicolumn{6}{c}{\textbf{VerilogEval v2}} \\
\cmidrule(lr){3-8}
& & \multicolumn{3}{c}{\textbf{Syntax}} & \multicolumn{3}{c}{\textbf{Function}} \\
\cmidrule(lr){3-5} \cmidrule(lr){6-8} 
\multirow{-4}{*}{\textbf{Stage}} & \multirow{-4}{*}{\textbf{Model}} &
\p${1}$ & \p${5}$ & \p${10}$ & \pass${1}$ & \pass${5}$ & \pass${10}$ \\
\midrule

& \textbf{RTLCoder} & 86.9 & 95.6 & 97.2 & 39.9 & 47.9 & 50.3 \\
& \textbf{CodeV} & 95.4 & 96.8 & 97.5 & 47.1 & 54.4 & 56.2 \\
& \textbf{HaVen} & 97.1 & 98.9 & 98.9 & 47.2 & 52.3 & 53.6 \\ 
\multirow{-4}{*}{\textbf{SFT}} & \textbf{OriGen} & 94.9 & 96.7 & 96.8 & 51.3 & 56.3 & 58.0 \\ \midrule

& \textbf{RTLCoder} & 86.4 & 97.0 & 97.9 & 39.6\minus{0.3} & 50.3\plus{2.4} & 53.0\plus{2.7} \\
& \textbf{CodeV} & 96.0 & 97.0 & 97.2 & 47.4\plus{0.3} & 54.9\plus{0.5} & 56.8\plus{0.6} \\
& \textbf{HaVen} & 94.4 & 99.0 & 99.2 & 48.2\plus{1.0} & 52.6\plus{0.3} & 53.6\plus{0.0} \\
\multirow{-4}{*}{\textbf{DPO}} & \textbf{OriGen} & 96.2 & 97.4 & 97.4 & 52.4\plus{1.1} & 56.7\plus{0.4} & 58.1\plus{0.1} \\ \bottomrule

\end{tabular}
\begin{tablenotes} \footnotesize
\item[+] All models are based on DeepSeek-Coder with 6.7B parameters.
\end{tablenotes}
\end{threeparttable}
\end{table}

%% file: tables/rtllm2_general.tex
\begin{table}[t]
\centering
\caption{Aligned Verilog Generation LLMs on RTLLM v2.}
\label{tab:rtllm2_general}
\begin{threeparttable}
\setlength{\tabcolsep}{2pt}
\footnotesize
\begin{tabular}{c c ccc ccc}
\toprule
& & \multicolumn{6}{c}{\textbf{RTLLM v2}} \\
\cmidrule(lr){3-8}
& & \multicolumn{3}{c}{\textbf{Syntax}} & \multicolumn{3}{c}{\textbf{Function}} \\
\cmidrule(lr){3-5} \cmidrule(lr){6-8} 
\multirow{-3}{*}{\textbf{Stage}} & \multirow{-3}{*}{\textbf{Model}} &
\p${1}$ & \p${5}$ & \p${10}$ & \pass${1}$ & \pass${5}$ & \pass${10}$ \\
\midrule

& \textbf{RTLCoder} & 82.9 & 90.1 & 92.3 & 43.5 & 48.0 & 50.8 \\
& \textbf{CodeV} & 75.5 & 90.7 & 94.9 & 45.5 & 55.7 & 60.9 \\
& \textbf{HaVen} & 89.0 & 94.2 & 95.7 & 50.6 & 60.5 & 64.8 \\ 
\multirow{-4}{*}{\textbf{SFT}} & \textbf{OriGen} & 77.5 & 87.8 & 89.5 & 40.9 & 57.1 & 61.9 \\ \midrule

& \textbf{RTLCoder} & 82.7 & 89.6 & 91.9 & 44.0\plus{0.5} & 49.8\plus{1.8} & 51.0\plus{0.2} \\
& \textbf{CodeV} & 78.1 & 91.4 & 95.5 & 47.7\plus{2.2} & 57.1\plus{1.4} & 61.4\plus{0.5} \\
& \textbf{HaVen} & 87.9 & 93.1 & 94.5 & 52.4\plus{1.8} & 62.1\plus{1.6} & 65.7\plus{0.9} \\
\multirow{-4}{*}{\textbf{DPO}} & \textbf{OriGen} & 76.5 & 89.9 & 91.9 & 42.3\plus{1.4} & 59.4\plus{2.3} & 64.0\plus{2.1} \\ \bottomrule

\end{tabular}
\begin{tablenotes} \footnotesize
\item[+] All models are based on DeepSeek-Coder with 6.7B parameters.
\end{tablenotes}
\end{threeparttable}
\vspace{-1em}
\end{table}

%% file: figs/simple_spec.tex
\begin{figure}[h]
\centering
\begin{tcolorbox}[left=1mm, right=1mm, top=0.5mm, bottom=0.5mm,  colback=white, colbacktitle=blue, colframe=black, fonttitle=\bfseries, width=0.96\linewidth]
\begin{lstlisting}[style=plain]
The Verilog code implements a serial audio decoder that processes audio data received over serial input (SDIN) based on clock signals (SCLK and LRCLK). It decodes audio samples of various widths (up to 32 bits) in either I2S or PCM formats, detects channel changes, and outputs the decoded audio along with validity and error status. Key functionalities include shifting in audio samples, determining the current audio channel (left/right), and handling errors based on the expected sample size. The output signals indicate if the decoded audio is valid and whether it's from the left channel or not.
\end{lstlisting}

\end{tcolorbox}
\caption{Simple design specification example.}
\label{fig:simple_spec}
\end{figure}

%% file: tables/ablation.tex
\begin{table}[t]
\centering
\caption{Ablation Study.}
\label{tab:ablation}
\begin{threeparttable}
\setlength{\tabcolsep}{2.5pt}
\begin{tabular}{cccccccccc}
\toprule
\multirow{4}{*}{\textbf{Stage}} & \multirow{4}{*}{\textbf{Method}} & \multirow{4}{*}{\textbf{\#Data}} & \multicolumn{3}{c}{\textbf{RTLLM v2}} & \multicolumn{3}{c}{\textbf{VerilogEval v2}} \\
\cmidrule(lr){4-6} \cmidrule(lr){7-9}
& & & \multicolumn{3}{c}{\textbf{Function}} & \multicolumn{3}{c}{\textbf{Function}} \\
\cmidrule(lr){4-6} \cmidrule(lr){7-9}
& & & \p${1}$ & \p${5}$ & \p${10}$ & \p${1}$ & \p${5}$ & \p${10}$ \\
\midrule
\multirow{3}{*}{\textbf{SFT}}
 & \oursft & 86,672 & 49.8 & 62.5 & 66.1 & 47.8 & 64.3 & 69.1 \\
 & Simple & 86,672 & 47.5 & 58.8 & 61.3 & 45.1 & 60.8 & 66.8 \\
 & Verification Insights & 1,796 & 49.7 & 61.9 & 65.4 & 48.0 & 64.8 & 69.0 \\
\midrule

\multirow{4}{*}{\textbf{PPO}}
& {AST} & 4,222 & 47.5 & 58.8 & 61.2 & 47.0 & 59.1 & 63.1 \\
& {DFG} & 2,834 & 48.4 & 57.9 & 61.9 & 45.5 & 60.9 & 62.7 \\
& {BLEU} & 4,706 & 48.0 & 57.5 & 63.2 & 46.2 & 60.5 & 61.4 \\
& {Testbench} & 1,796 & 49.3 & 58.4 & 62.1 & 45.4 & 59.4 & 62.1 \\
\midrule

\multirow{5}{*}{\textbf{DPO}}
& {AST} & 4,222 & 49.2 & 60.5 & 63.1 & 48.7 & 62.0 & 66.1 \\
& {DFG} & 2,834 & 51.5 & 62.6 & 66.3 & 48.1 & 61.7 & 66.1 \\
& {BLEU} & 4,706 & \cellcolor{gold}52.6 & 62.0 & 66.6 & 48.2 & 61.2 & 65.3 \\
& {Testbench w/ fails} & 3,560 & 49.3 & 59.5 & 64.0 & 47.6 & 58.3 & 64.6 \\
& {\oursrl} & 1,796 & 52.4 & \cellcolor{gold}{66.4} & \cellcolor{gold}{69.0} & \cellcolor{gold}{49.3} & \cellcolor{gold}{64.7} & \cellcolor{gold}{69.5} \\

\bottomrule
\end{tabular}
\begin{tablenotes}\footnotesize
\item[+] The background colors \colorbox{gold}{Gold} represents the best performance.
\item[*] All experiments are based on \oursft-Qwen2.5-Coder.
\end{tablenotes}
\end{threeparttable}
\vspace{-1em}
\end{table}

%% file: figs/ppl.tex
\begin{figure}[t]
    \centering
    \includegraphics[width=0.8\linewidth]{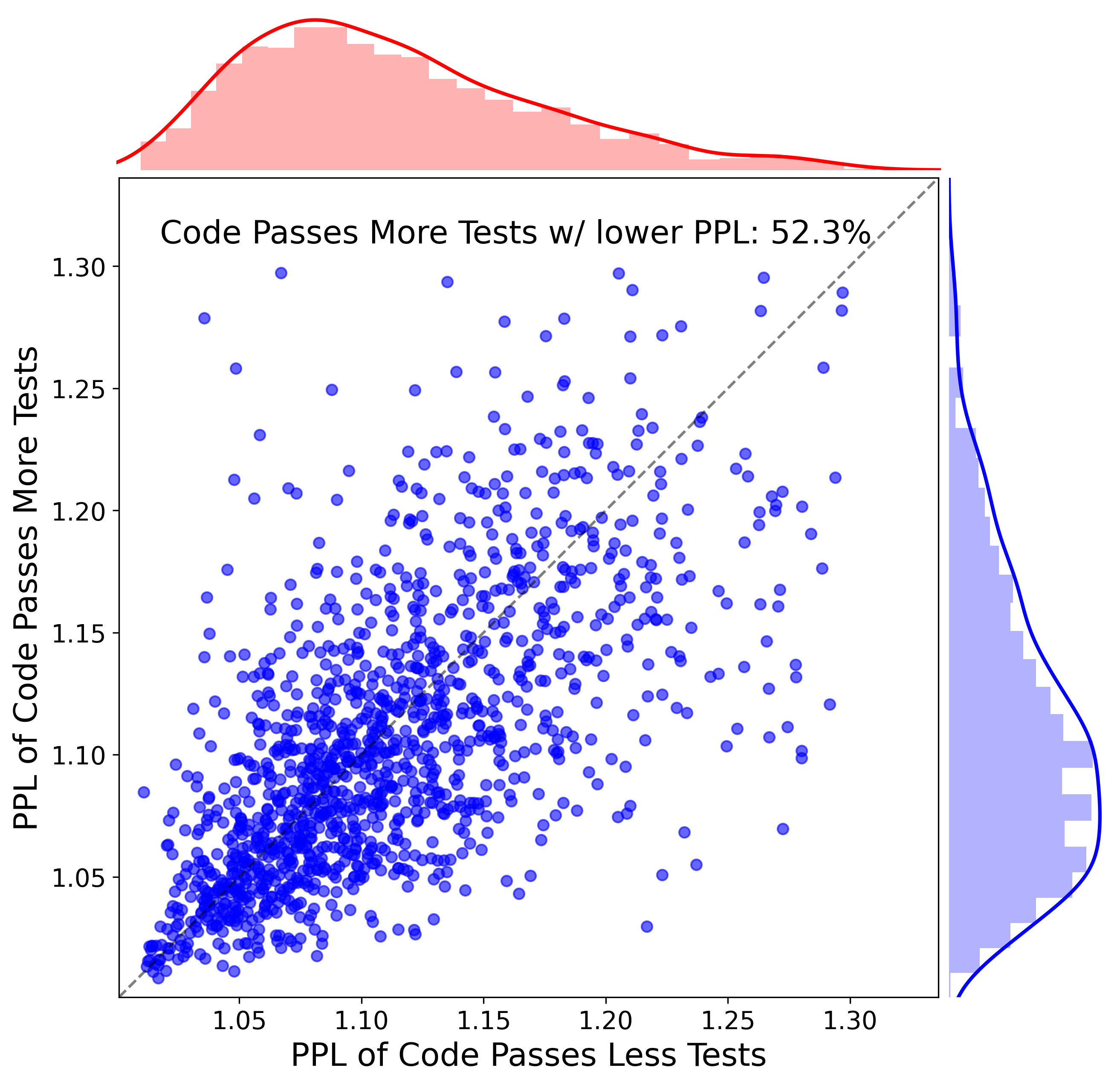}
    \caption{The PPL of code pairs generated by the fine-tuned model.}
    \label{fig:misalign}
    \vspace{-2em}
\end{figure}

%% file: prompts/spec.tex
\begin{tcolorbox}[title=Generate Specification, left=0mm, right=0mm, top=0mm, bottom=0mm,  colback=white, fonttitle=\footnotesize]
\begin{lstlisting}[style=plain]
Your current task is to provide design specifications for the RTL module code. Please carefully analyze the logic of the code to avoid easily determining its function based on signal names or naming conventions.
Use specific signal names instead of using the meaning of the signal as a substitute.
You need to analyze the design concept of this module and explain why it is designed in this way. The goal is to empower other design engineers to accurately implement the corresponding functions.
Please output in the following format:
Module name:
  - The name of the module.
Module functions
  - Describe the primary function and purpose of the module IN DETAIL. List each primary function and describe it.
Input and output port descriptions
  - Describe all ports, including their names, widths, data types, purposes. Highlight specific ones like clock (CLK) or reset (RST). 
  ...
Internal Working Principle
  - Explain the high-level operation of the module IN DETAIL, and describe how inputs are processed to generate outputs. 
Implementation Logic Explanation
  - Provide a DETAILED description of the module implementation logic, such as, but not limited to: basic operation principle, output behavior, critical combinational logic, etc.
Internally Defined Signal Descriptions
  - List and explain important internal signals, their purposes and relationships with other signals.
The description ends with a brief summary of the module.
This is the Verilog module: (*@\colorbox{keybackgreen}{\textbf{[Code]}}@*)
\end{lstlisting}
\end{tcolorbox}

%% file: prompts/improve.tex
\begin{tcolorbox}[title=Improve Testbench, left=0mm, right=0mm, top=0mm, bottom=0mm,  colback=white, fonttitle=\footnotesize]

\begin{lstlisting}[style=plain]
You have successfully generated a testbench, but the line coverage is below the threshold of 90%. The detailed line coverage report is provided below: (*@\colorbox{silver}{\textbf{[Report]}}@*)
\end{lstlisting}
\begin{lstlisting}[style=plain, aboveskip=2pt]
In this report, lines are marked as either 0/1, indicating that the line is not covered, or 1/1, indicating that the line is covered. Your objective is to create new testcases that specifically target the lines marked as 0/1 to enhance the overall coverage. You can disregard Verilog keywords like "if," "else".
\end{lstlisting}
\end{tcolorbox}

%% file: prompts/funcpoint.tex
\begin{tcolorbox}[title=Generate Function Points, left=0mm, right=0mm, top=0mm, bottom=0mm,  colback=white, fonttitle=\footnotesize]
\begin{lstlisting}[style=plain]
You are a digital integrated circuit (IC) design engineer. Please analyze the functional points of the module in detail according to the content of the module specification document you have. In your analysis, consider various typical application scenarios like:
  - Input port incentive scenarios
  - Supported working modes
  ...

Your task is to return the top three most important functional points, presented in the following JSON format:
{
  1: {
      "Point": "[Function Point Category]",
      "Scenario": "[Detailed description of the scenario]",
      "Application": "[Use case for this functionality]"
    },
  ...
}

You only need to return the top-3 most important functional points.
This is the specification: (*@\colorbox{keybackblue}{\textbf{[Specification]}}@*)
\end{lstlisting}
\end{tcolorbox}

%% file: prompts/testcase.tex
\begin{tcolorbox}[title=Generate Test Cases, left=0mm, right=0mm, top=0mm, bottom=0mm,  colback=white, fonttitle=\footnotesize]
\begin{lstlisting}[style=plain]
Now you need to systematically disassemble the functionality and create a series of detailed test cases based on function points to ensure comprehensive validation across all dimensions.

You only need to return the top five most important testcases. Return the test cases in JSON format with the following structure:
{
  1: {
    "Title": "[Descriptive name of the test case]",
    "Objective": "[Clear statement of what to verify]",
    "Setup": "[Step-by-step description of how to configure]",
    "Coverage": "[Detailed explanation of the function points and validation dimensions covered by this test]"
  },
  ...
}

You only need to return the top-5 most important testcases.
\end{lstlisting}
\end{tcolorbox}

%% file: prompts/draft.tex
\begin{tcolorbox}[title=Draft Testbench, left=1mm, right=1mm, top=0.5mm, bottom=0.5mm,  colback=white, fonttitle=\footnotesize]
\begin{lstlisting}[style=plain]
You are now required to create a Verilog testbench. The testbench must instantiate the DUT with all ports properly connected and generate appropriate clock and reset signals. It should include comprehensive test vectors that cover normal operation, edge cases and error conditions. The final code must be self-contained, ready for simulation, and conform to Verilog coding practices.

The testbench only needs to display the DUT's output under the conditions set by the test cases. All test cases must be tested, and none should be ignored or omitted. You need to return the testbench with the following template:
\end{lstlisting}
\vspace{-2mm}
\begin{lstlisting}[style=verilog]
`timescale 1ns / 1ps
module testbench;
// Parameter or localparam ...
// Input, output or input ports ...
// Instantiate the Design Under Test (DUT) ...
// Clock toggle ...

initial begin
  // Initialize Inputs ...
  // Wait 100 ns for global reset to finish ...
  // Must set stimulus and report all testcases with simulation results
  $display("===========TestCases===========")
  // Test Case 1:
  $display("Test Case 1. Expected x1: [VALUE], ...")
  $display("Test Case 1. Actual x1: \%d, ...", x1, ...)
  ...
  $display("===========End===========")
  // Must finish simulation
  $finish;
end
endmodule
\end{lstlisting}
\begin{lstlisting}[style=plain]

Below is the Design Under Test (DUT): (*@\colorbox{keybackgreen}{\textbf{[Code]}}@*)
\end{lstlisting}
\end{tcolorbox}

%% file: prompts/rectify.tex
\begin{tcolorbox}[title=Rectify Testbench, left=0mm, right=0mm, top=0mm, bottom=0mm,  colback=white, fonttitle=\footnotesize]
\begin{lstlisting}[style=plain]
You have designed a testbench setting according to the specification and testcases for golden dut. You must report the final test result. You should record the error_count if the dut responses do not match the testbench.

You should add the test score part:
\end{lstlisting}
\vspace{-2mm}
\begin{lstlisting}[style=verilog]
  // same as before ...
  // Be sure to report the final test score
  if (error_count == 0) begin
      $display("Your Design Passed");
    end
  else begin
    $display("Test with %d failures", error_count);
  end
  // Must finish simulation
  $finish;
end
endmodule
\end{lstlisting}
\begin{lstlisting}[style=plain]
You have already simulate the dut and got the actual output. You should modify the Expected result in the testbench to match the actual output. 

Below is the simulation output: (*@\colorbox{silver}{\textbf{[Simulation]}}@*)
\end{lstlisting}
\end{tcolorbox}

%% file: main.bbl
\begin{thebibliography}{10}

\bibitem{rewardhacking}
D.~Amodei, C.~Olah, J.~Steinhardt, P.~Christiano, J.~Schulman, and D.~Man{\'e}.
\newblock Concrete problems in ai safety.
\newblock {\em arXiv preprint arXiv:1606.06565}, 2016.

\bibitem{claude}
Anthropic.
\newblock Claude 3 sonnet.
\newblock Large language model, 2024.
\newblock Accessed: April 15, 2025.

\bibitem{codeqwen}
J.~Bai, S.~Bai, Y.~Chu, Z.~Cui, K.~Dang, X.~Deng, Y.~Fan, W.~Ge, Y.~Han, F.~Huang, et~al.
\newblock Qwen technical report.
\newblock {\em arXiv preprint arXiv:2309.16609}, 2023.

\bibitem{rlhf}
Y.~Bai, A.~Jones, K.~Ndousse, A.~Askell, A.~Chen, N.~DasSarma, D.~Drain, S.~Fort, D.~Ganguli, T.~Henighan, et~al.
\newblock Training a helpful and harmless assistant with reinforcement learning from human feedback.
\newblock {\em arXiv preprint arXiv:2204.05862}, 2022.

\bibitem{passnk}
M.~Chen, J.~Tworek, H.~Jun, Q.~Yuan, H.~P. D.~O. Pinto, J.~Kaplan, H.~Edwards, Y.~Burda, N.~Joseph, G.~Brockman, et~al.
\newblock Evaluating large language models trained on code.
\newblock {\em arXiv preprint arXiv:2107.03374}, 2021.

\bibitem{origen}
F.~Cui, C.~Yin, K.~Zhou, Y.~Xiao, G.~Sun, Q.~Xu, Q.~Guo, D.~Song, D.~Lin, X.~Zhang, et~al.
\newblock Origen: Enhancing rtl code generation with code-to-code augmentation and self-reflection.
\newblock {\em arXiv preprint arXiv:2407.16237}, 2024.

\bibitem{autovcoder}
M.~Gao, J.~Zhao, Z.~Lin, W.~Ding, X.~Hou, Y.~Feng, C.~Li, and M.~Guo.
\newblock Autovcoder: A systematic framework for automated verilog code generation using llms.
\newblock In {\em 2024 IEEE 42nd International Conference on Computer Design (ICCD)}, pages 162--169, 2024.

\bibitem{llama3}
A.~Grattafiori, A.~Dubey, A.~Jauhri, A.~Pandey, A.~Kadian, A.~Al-Dahle, A.~Letman, A.~Mathur, A.~Schelten, A.~Vaughan, et~al.
\newblock The llama 3 herd of models.
\newblock {\em arXiv preprint arXiv:2407.21783}, 2024.

\bibitem{deepseekcoder_v1.5}
D.~Guo, Q.~Zhu, D.~Yang, Z.~Xie, K.~Dong, W.~Zhang, G.~Chen, X.~Bi, Y.~Wu, Y.~Li, et~al.
\newblock Deepseek-coder: When the large language model meets programming--the rise of code intelligence.
\newblock {\em arXiv preprint arXiv:2401.14196}, 2024.

\bibitem{verilogcoder}
C.-T. Ho, H.~Ren, and B.~Khailany.
\newblock Verilogcoder: Autonomous verilog coding agents with graph-based planning and abstract syntax tree (ast)-based waveform tracing tool.
\newblock In {\em Proceedings of the AAAI Conference on Artificial Intelligence}, volume~39, pages 300--307, 2025.

\bibitem{codeqwen2}
B.~Hui, J.~Yang, Z.~Cui, J.~Yang, D.~Liu, L.~Zhang, T.~Liu, J.~Zhang, B.~Yu, K.~Lu, et~al.
\newblock Qwen2. 5-coder technical report.
\newblock {\em arXiv preprint arXiv:2409.12186}, 2024.

\bibitem{gpt4o}
A.~Hurst, A.~Lerer, A.~P. Goucher, A.~Perelman, A.~Ramesh, A.~Clark, A.~Ostrow, A.~Welihinda, A.~Hayes, A.~Radford, et~al.
\newblock Gpt-4o system card.
\newblock {\em arXiv preprint arXiv:2410.21276}, 2024.

\bibitem{mistral}
A.~Q. Jiang, A.~Sablayrolles, A.~Mensch, C.~Bamford, D.~S. Chaplot, D.~de~las Casas, F.~Bressand, G.~Lengyel, G.~Lample, L.~Saulnier, L.~R. Lavaud, M.-A. Lachaux, P.~Stock, T.~L. Scao, T.~Lavril, T.~Wang, T.~Lacroix, and W.~E. Sayed.
\newblock Mistral 7b, 2023.

\bibitem{decompose}
T.~Khot, H.~Trivedi, M.~Finlayson, Y.~Fu, K.~Richardson, P.~Clark, and A.~Sabharwal.
\newblock Decomposed prompting: A modular approach for solving complex tasks.
\newblock {\em arXiv preprint arXiv:2210.02406}, 2022.

\bibitem{ircoco}
B.~Li, Z.~Sun, T.~Huang, H.~Zhang, Y.~Wan, G.~Li, Z.~Jin, and C.~Lyu.
\newblock Ircoco: Immediate rewards-guided deep reinforcement learning for code completion.
\newblock {\em Proceedings of the ACM on Software Engineering}, 1(FSE):182--203, 2024.

\bibitem{drattack}
X.~Li, R.~Wang, M.~Cheng, T.~Zhou, and C.-J. Hsieh.
\newblock Drattack: Prompt decomposition and reconstruction makes powerful llm jailbreakers.
\newblock {\em arXiv preprint arXiv:2402.16914}, 2024.

\bibitem{deepseekv3}
A.~Liu, B.~Feng, B.~Xue, B.~Wang, B.~Wu, C.~Lu, C.~Zhao, C.~Deng, C.~Zhang, C.~Ruan, et~al.
\newblock Deepseek-v3 technical report.
\newblock {\em arXiv preprint arXiv:2412.19437}, 2024.

\bibitem{verilogeval}
M.~Liu, N.~Pinckney, B.~Khailany, and H.~Ren.
\newblock Verilogeval: Evaluating large language models for verilog code generation.
\newblock In {\em 2023 IEEE/ACM International Conference on Computer Aided Design (ICCAD)}, pages 1--8. IEEE, 2023.

\bibitem{rtlcoder}
S.~Liu, W.~Fang, Y.~Lu, J.~Wang, Q.~Zhang, H.~Zhang, and Z.~Xie.
\newblock Rtlcoder: Fully open-source and efficient llm-assisted rtl code generation technique.
\newblock {\em IEEE Transactions on Computer-Aided Design of Integrated Circuits and Systems}, 2024.

\bibitem{rtllm2}
S.~Liu, Y.~Lu, W.~Fang, M.~Li, and Z.~Xie.
\newblock Openllm-rtl: Open dataset and benchmark for llm-aided design rtl generation.
\newblock {\em arXiv preprint arXiv:2503.15112}, 2025.

\bibitem{reject_sampling}
T.~Liu, Y.~Zhao, R.~Joshi, M.~Khalman, M.~Saleh, P.~J. Liu, and J.~Liu.
\newblock Statistical rejection sampling improves preference optimization.
\newblock {\em arXiv preprint arXiv:2309.06657}, 2023.

\bibitem{rtllm}
Y.~Lu, S.~Liu, Q.~Zhang, and Z.~Xie.
\newblock Rtllm: An open-source benchmark for design rtl generation with large language model.
\newblock In {\em 2024 29th Asia and South Pacific Design Automation Conference (ASP-DAC)}, pages 722--727. IEEE, 2024.

\bibitem{pyranet}
B.~Nadimi, G.~O. Boutaib, and H.~Zheng.
\newblock Pyranet: A multi-layered hierarchical dataset for verilog.
\newblock {\em arXiv preprint arXiv:2412.06947}, 2024.

\bibitem{bleu}
K.~Papineni, S.~Roukos, T.~Ward, and W.-J. Zhu.
\newblock Bleu: a method for automatic evaluation of machine translation.
\newblock In {\em Proceedings of the 40th annual meeting of the Association for Computational Linguistics}, pages 311--318, 2002.

\bibitem{verilogeval2}
N.~Pinckney, C.~Batten, M.~Liu, H.~Ren, and B.~Khailany.
\newblock Revisiting verilogeval: A year of improvements in large-language models for hardware code generation.
\newblock {\em ACM Transactions on Design Automation of Electronic Systems}, 2025.

\bibitem{tool}
Y.~Qin, S.~Hu, Y.~Lin, W.~Chen, N.~Ding, G.~Cui, Z.~Zeng, X.~Zhou, Y.~Huang, C.~Xiao, et~al.
\newblock Tool learning with foundation models.
\newblock {\em ACM Computing Surveys}, 57(4):1--40, 2024.

\bibitem{dpo}
R.~Rafailov, A.~Sharma, E.~Mitchell, C.~D. Manning, S.~Ermon, and C.~Finn.
\newblock Direct preference optimization: Your language model is secretly a reward model.
\newblock {\em Advances in Neural Information Processing Systems}, 36:53728--53741, 2023.

\bibitem{hallucination}
V.~Rawte, A.~Sheth, and A.~Das.
\newblock A survey of hallucination in large foundation models.
\newblock {\em arXiv preprint arXiv:2309.05922}, 2023.

\bibitem{codellama}
B.~Roziere, J.~Gehring, F.~Gloeckle, S.~Sootla, I.~Gat, X.~E. Tan, Y.~Adi, J.~Liu, R.~Sauvestre, T.~Remez, et~al.
\newblock Code llama: Open foundation models for code.
\newblock {\em arXiv preprint arXiv:2308.12950}, 2023.

\bibitem{ppocoder}
P.~Shojaee, A.~Jain, S.~Tipirneni, and C.~K. Reddy.
\newblock Execution-based code generation using deep reinforcement learning.
\newblock {\em arXiv preprint arXiv:2301.13816}, 2023.

\bibitem{pyverilog}
S.~Takamaeda-Yamazaki.
\newblock Pyverilog: A python-based hardware design processing toolkit for verilog hdl.
\newblock In {\em Applied Reconfigurable Computing: 11th International Symposium, ARC 2015, Bochum, Germany, April 13-17, 2015, Proceedings 11}, pages 451--460. Springer, 2015.

\bibitem{verigen}
S.~Thakur, B.~Ahmad, H.~Pearce, B.~Tan, B.~Dolan-Gavitt, R.~Karri, and S.~Garg.
\newblock Verigen: A large language model for verilog code generation.
\newblock {\em ACM Transactions on Design Automation of Electronic Systems}, 29(3):1--31, 2024.

\bibitem{veriseek}
N.~Wang, B.~Yao, J.~Zhou, X.~Wang, Z.~Jiang, and N.~Guan.
\newblock Large language model for verilog generation with golden code feedback.
\newblock {\em arXiv preprint arXiv:2407.18271}, 2024.

\bibitem{haven}
Y.~Yang, F.~Teng, P.~Liu, M.~Qi, C.~Lv, J.~Li, X.~Zhang, and Z.~He.
\newblock Haven: Hallucination-mitigated llm for verilog code generation aligned with hdl engineers.
\newblock In {\em Design, Automation \& Test in Europe (DATE)}, 2025.

\bibitem{react}
S.~Yao, J.~Zhao, D.~Yu, N.~Du, I.~Shafran, K.~Narasimhan, and Y.~Cao.
\newblock React: Synergizing reasoning and acting in language models.
\newblock In {\em International Conference on Learning Representations (ICLR)}, 2023.

\bibitem{betterv}
P.~Zehua, H.~Zhen, M.~Yuan, Y.~Huang, and B.~Yu.
\newblock Betterv: Controlled verilog generation with discriminative guidance.
\newblock In {\em Forty-first International Conference on Machine Learning}, 2024.

\bibitem{plum}
D.~Zhang, S.~Diao, X.~Zou, and H.~Peng.
\newblock Plum: Improving code lms with execution-guided on-policy preference learning driven by synthetic test cases.
\newblock {\em arXiv preprint arXiv:2406.06887}, 2024.

\bibitem{flow}
Z.~Zhang, Y.~Sun, J.~Ye, T.-S. Liu, J.~Zhang, and Y.~Yu.
\newblock Flow to better: Offline preference-based reinforcement learning via preferred trajectory generation.
\newblock In {\em The Twelfth International Conference on Learning Representations}, 2023.

\bibitem{codev}
Y.~Zhao, D.~Huang, C.~Li, P.~Jin, Z.~Nan, T.~Ma, L.~Qi, Y.~Pan, Z.~Zhang, R.~Zhang, et~al.
\newblock Codev: Empowering llms for verilog generation through multi-level summarization.
\newblock {\em arXiv preprint arXiv:2407.10424}, 2024.

\bibitem{lesat2most}
D.~Zhou, N.~Sch{\"a}rli, L.~Hou, J.~Wei, N.~Scales, X.~Wang, D.~Schuurmans, C.~Cui, O.~Bousquet, Q.~Le, et~al.
\newblock Least-to-most prompting enables complex reasoning in large language models.
\newblock {\em arXiv preprint arXiv:2205.10625}, 2022.

\end{thebibliography}
